\documentclass[12pt,preprint]{aastex}

\shorttitle{Spectral Variability of NGC 4151} \shortauthors{Wang et al.}

\begin{document}

\title{Revisit Short Term X-ray Spectral Variability of NGC 4151 with {\em Chandra}}


\author{Junfeng Wang,\altaffilmark{1} G. Risaliti,\altaffilmark{1,2} G. Fabbiano,\altaffilmark{1} M. Elvis,\altaffilmark{1} A. Zezas,\altaffilmark{1} and M. Karovska\altaffilmark{1}}

\altaffiltext{1}{Harvard-Smithsonian Center for Astrophysics, 60
  Garden St, Cambridge, MA 02138} \email{juwang@cfa.harvard.edu;
  risaliti@cfa.harvard.edu; pepi@cfa.harvard.edu;
  elvis@cfa.harvard.edu; azezas@cfa.harvard.edu; karovska@cfa.harvard.edu}

\altaffiltext{2}{Current Address: INAF-Arcetri Observatory, Largo E,
  Fermi 5, I-50125 Firenze, Italy}

\begin{abstract}

We present new X-ray spectral data for the Seyfert 1 nucleus in NGC
4151 observed with {\em Chandra} for $\sim$200 ks.  A significant ACIS
pileup is present, resulting in a non-linear count rate variation
during the observation.  With pileup corrected spectral fitting, we
are able to recover the spectral parameters and find consistency with
those derived from unpiled events in the ACIS readout streak and outer
region from the bright nucleus.  The absorption corrected 2--10 keV
flux of the nucleus varied between $6\times 10^{-11}$ erg s$^{-1}$
cm$^{-2}$ and $10^{-10}$ erg s$^{-1}$ cm$^{-2}$ ($L_{2-10 {\rm
    keV}}\sim 1.3-2.1\times 10^{42}$ erg s$^{-1}$). Similar to earlier
$Chandra$ studies of NGC 4151 at a historical low state, the photon
indices derived from the same absorbed power-law model are $\Gamma\sim
0.7-0.9$.  However, we show that $\Gamma$ is highly dependent on the
adopted spectral models.  Fitting the power-law continuum with a
Compton reflection component gives $\Gamma\sim 1.1$.  By including
passage of non-uniform X-ray obscuring clouds, we can reproduce the
apparent flat spectral states with $\Gamma\sim 1.7$, typical for
Seyfert 1 AGNs.  The same model also fits the hard spectra from
previous {\em ASCA} ``long look'' observation of NGC 4151 in the
lowest flux state.  The spectral variability during our observation
can be interpreted as variations in intrinsic soft continuum flux
relative to a Compton reflection component that is from distant cold
material and constant on short time scale, or variations of partially
covering absorber in the line of sight towards the nucleus.  An
ionized absorber model with ionization parameter $\log \xi\sim
0.8-1.1$ can also fit the low-resolution ACIS spectra.  If the partial
covering model is correct, adopting a black hole mass $M_{BH}\sim
4.6\times 10^7$M$_{\odot}$ we constrain the distance of the obscuring
cloud from the central black hole to be $r\lesssim 9$ light-days,
consistent with the size of broad emission line region of NGC 4151
from optical reverberation mapping.

\end{abstract}


\section{Introduction}\label{intro}

%4piD^2=2.7E+52

NGC 4151 is a well-known nearby ($D\sim 13.3$ Mpc for $H_0=75$ km
s$^{-1}$ Mpc$^{-1}$; Mundell et al. 1999) bright Seyfert 1.5 galaxy
(Osterbrock \& Koski 1976), hosting one of the apparently brightest
active galactic nuclei (AGN) (Crenshaw \& Kraemer 2007;
Storchi-Bergmann et al. 2009). For a review, Ulrich (2000) provides
the multiwavelength properties of the galaxy, and the ultraviolet
(UV)/X-ray spectra of the AGN.  As an archetype of its class, the NGC
4151 nucleus has been intensively studied with all the major X-ray
observatories. Its broad band X-ray spectrum is very complex; the main
features are summarized as follows:

(1) The 2--10 keV (absorption corrected) luminosity is significantly
variable, in the range $\sim 2-20\times 10^{42}$ erg s$^{-1}$
($F_{2-10 keV}\sim 5-50\times 10^{-11}$ erg s$^{-1}$) with hard spectrum
($>4$ keV) characterized by a power law (Yaqoob et al. 1993; Warwick
et al. 1995, 1996). The 2--10 keV flux can double on timescale of
$\sim 0.5$ days (Tananbaum et al. 1978; Yaqoob \& Warwick 1991) and
flaring events are seen on timescales of days and weeks (Elvis 1976;
Lawrence 1980; Edelson et al. 1996; Markowitz et al. 2003; de Rosa et
al. 2007).

(2) The photon spectral index ($\Gamma$) of the hard power law
spectrum varies between $\sim 1.35$ and $\sim 1.7$, and correlates
with the 2--10 keV flux\footnote{The linear relation between $\Gamma$
  and the absorption corrected 2--10 keV flux in units of $10^{-11}$
  erg s$^{-1}$ derived from EXOSAT and Ginga measurements is
  $\Gamma=1.18+0.012F_{2-10keV}^c$ (Perola et al. 1986) and
  $\Gamma=1.35+0.011F_{2-10keV}^c$ (Yaqoob \& Warwick 1991),
  respectively.}. The spectrum becomes softer with increasing flux,
and the same variability in the X-ray continuum flux is observed on
time scales of days to years in the spectral slope. Note that the
range of $\Gamma$ corresponds to a harder spectrum than the canonical
value for Seyferts of $\Gamma\sim 1.8-1.9$ (e.g., Mushotzky 1984,
Nandra \& Pounds 1994).

(3) A soft ``excess'' ($<2$ keV) is present in the spectra when fitted
with the power law measured in the hard-band and a uniform absorption
(Holt et al. 1980). Extended soft X-ray emission associated with
ionized gas observed in the optical has been detected (Elvis et
al. 1983; Morse et al. 1995; Ogle et al. 2000; Yang et al 2001). The
soft flux shows stability against the high variations in the UV and
hard X-ray continua (Perola et al. 1986; Weaver et al. 1994a,b).

(4) A clear emission line is present at $6.39\pm 0.07$ keV (Matsuoka
et al. 1986; Wang et al. 2001; Schurch et al. 2003), consistent with
being Fe$K_{\alpha}$ emission line produced by the fluoresence of cold
iron illuminated by the X-ray continuum. The line has a relatively
narrow Gaussian profile and the line flux remains constant over time
(Warwick et al. 1989; Schurch et al. 2003). Long term monitoring with
{\em RXTE}/PCA over 800 days (Markowitz et al. 2003) shows a decreasing line
flux. De Rosa et al. (2007) suggest that the line exhibits some
variability related to the reflection component.

Particularly interesting results were reported from previous {\em
  Chandra} observations. Yang et al. (2001) presented {\em Chandra}
observations of NGC 4151 and found that the 2--9 keV spectrum of the
nucleus is described by a heavily absorbed ($N_H\simeq 3\times
10^{22}$ cm$^{-2}$), extremely hard power law (photon index $\Gamma=
0.32^{+0.05}_{-0.12}$). This is consistent with the {\em Chandra} HETG
spectra reported in Ogle et al. (2000), in which the hard continuum
emission is characterized by a photon index $\Gamma=0.4\pm 0.3$ with a
column of $N_H\simeq 3.7\times 10^{22}$ cm$^{-2}$.  Note that both
observations found NGC 4151 in a low state with $F_{2-10keV} \sim
5.5\times 10^{-11}$ erg s$^{-1}$ and do not follow the
$\Gamma$--$F_{2-10keV}$ correlation.  The rather unusually flat power
law slope complicates the understanding of the X-ray spectrum of NGC
4151.

We have obtained deep {\em Chandra} observation aimed to take
advantage of {\em Chandra}'s sub-arcsecond spatial resolution (van
Speybroeck et al. 1997) to study the soft circum-nuclear extended
emission. These data also present an opportunity to examine the
spectra of this puzzling nucleus.  The {\em Chandra} observations and
data reduction are briefly described in \S~\ref{obs}.  We examine the
light curves of the nucleus and perform diagnostic analysis of the
X-ray spectra in \S~\ref{pileup_analysis}. In \S~\ref{models} we
explain the various ways to model the X-ray spectra.  We then analyzed
the archival deep {\em ASCA} observation of NGC 4151 in low state to check
the validity of our models (\S~\ref{asca_look}). Finally we discuss the
results in \S~\ref{discussion} and summarize our findings in
\S~\ref{summary}.

\section{Observations and Data Reduction}\label{obs}

NGC 4151 was imaged with the {\it Chandra X-ray Observatory}
(Weisskopf et al. 2002) on 2008 March 27 (ObsID 9218; 69 ks) and March
29 (ObsID 9217; 125 ks). Both observations were obtained with the
back-illuminated chip of the Advanced CCD Imaging Spectrometer
spectroscopy array (ACIS-S; Garmire et al. 2003) in Faint ``1/8
subarray'' mode.  CCDs S2, S3, and S4 were read out.  The NGC 4151
nucleus ($\alpha$=12$^h$10$^m$32.$^s$6,
$\delta$=+39$^{\circ}$24$^{\prime}$21$^{\prime\prime}$; Clements 1981)
was placed at the on-axis position near the ACIS-S3 aimpoint of {\em
  Chandra}'s High Resolution Mirror Assembly (HRMA; van Speybroeck et
al. 1997) at a 186$^{\circ}$ roll angle.  This reduced the frame time from
the nominal 3.2 s in full array mode to 0.6 s to reduce photon
pileup\footnote{For more information see {\em Chandra} ABC Guide to
  Pile Up, available at
  \url{http://cxc.harvard.edu/ciao/download/doc/pileup\_abc.ps}}--during
the exposure of a frame, two or more photons from a high count rate
X-ray source are recorded as a single event, causing loss of
information from the original events.

The data were processed following the standard {\em Chandra} ACIS data
preparation thread, with the {\it Chandra} X-Ray Center (CXC) CIAO
v4.0 software and HEASOFT v6.4 package\footnote{See
  \url{http://cxc.harvard.edu/ciao/} and
  \url{http://heasarc.gsfc.nasa.gov/lheasoft/} for more
  information.}. The
CALDB\footnote{\url{http://cxc.harvard.edu/caldb/}} v3.4.3 calibration
files were used.  To remove periods of high background, we extracted
light curves for chip S3 in both observations, excluding any bright
sources in the field.  Events during period of background flares were
screened out, and the resulting total exposure was 124 ks for ObsID
9217 and 67 ks for ObsID 9218 before dead time correction (after dead
time correction, 116 ks and 63 ks for ObsID 9217 and 9218,
respectively).

\begin{figure}[H]
\plotone{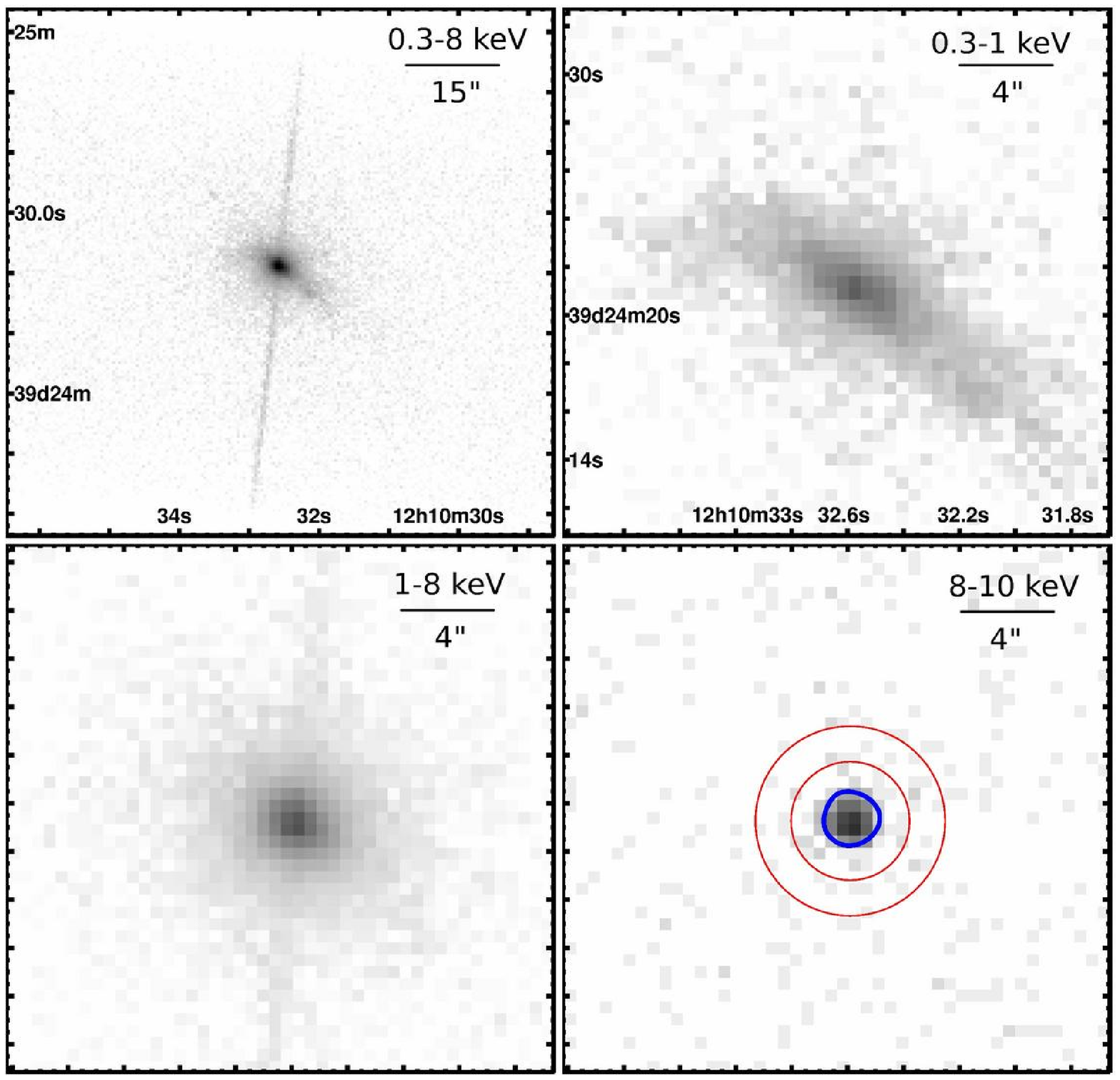}
\caption{(a) 0.3--8 keV ACIS image of the NGC 4151 galaxy. The narrow
  bright streak extending from both sides of the nucleus is not real
  source feature, but the ACIS CCD readout.  (b) Soft X-ray emission
  (0.3--1 keV) in the circumnuclear region.  The emission is extended
  along (c) Hard X-ray emission (1--8 keV) for the same field as
  (b). (d) 8--10 keV image of the same circumnuclear region
  demonstrating the high energy photons as a result of pile up.  The
  annulus outlines the wing region of the PSF used to evaluate the
  pileup (\S~\ref{specdiag}).
\label{image}}
\end{figure}

Figure~\ref{image} shows the ACIS view of the NGC 4151 galaxy and
images zoomed in to its nucleus region in different energy ranges
(0.3--1 keV, 1--8 keV, and 8--10 keV).  The X-ray emission comprises
an unresolved, bright nucleus and resolved soft extended regions
(Elvis et al. 1983; Morse et al. 1995; Yang et al. 2001) with spatial
scales of several hundred parsecs ($\sim 6\arcsec$ on both sides of
the nucleus).  Detailed ACIS study of the soft extended emission will
be reported in a separate paper, complemented by {\em Chandra}
High-Resolution Camera (HRC; Murray et al. 1997) data (Wang et
al. 2009).

Yang et al. (2001) reported that both their 0.4 s and 3.2 s frame time
ACIS images of the nucleus in low-flux state suffered considerable
photon pileup.  Even if NGC 4151 were at its lowest flux state, the
brightness of the nucleus would cause photon pileup in our sub-array
observations, despite the reduced frame time.  This is already evident
in Figure~\ref{image}: the nucleus remains bright in the hardest band,
while there is little effective area of HRMA/ACIS-S3 in the very hard
energy range ($E > 8$ keV; see the {\em Chandra} Proposers'
Observatory
Guide\footnote{\url{http://cxc.harvard.edu/proposer/POG/}}, POG). Most
of the hardest photons in the nucleus region actually come from piled
soft photons.  Pileup effect is a great concern to both timing and
spectral analysis.  Photon pileup can reduce the apparent count rate,
causing a non-linear relation between the source flux and the count
rate. It also skews the observed spectra towards higher energies.  The
impact of pileup to our data will be carefully evaluated in the next
section.

\section{Initial Analysis with Pileup}\label{pileup_analysis}

To have a better handle of the pileup and derive meaningful results,
we will examine the light curves and spectra of the bright nucleus,
using extracted data from the point spread function (PSF) core region,
the PSF wing region, and the ACIS readout streak\footnote{X-ray
  photons arrived during the frame transfer were still recorded but
  mispositioned along the axis of CCD readout. For a bright source,
  this results in a streak along the entire column of the source. See
  {\em Chandra} POG and McCollough \& Rots(2005).}.  By comparing the
analysis for the piled and unpiled data, we demonstrate that the
existing pileup modeling tool works well with our data.

\subsection{X-ray Light Curves}\label{ltcv}

Adopting the best-fit model parameters for NGC 4151 at low state
(e.g., Ogle et al. 2000, Yang et al. 2001) and a 0.6 s frame time, the
Portable Interactive Multi-Mission Simulator
(PIMMS\footnote{\url{http://heasarc.gsfc.nasa.gov/Tools/w3pimms.html}})
predicts a count rate of 1.4 count s$^{-1}$ (0.8 count s$^{-1}$ after
pileup) and a pileup fraction of $\sim$28\% (the estimated percentage
of detected events that consist of more than one photon; see the {\em
  Chandra} POG).  A higher observed flux of $2\times 10^{-10}$ erg
s$^{-1}$ cm$^{-2}$ (2-10 keV) was reported in Weaver et
al. (1994b). If the nucleus is in a similar high flux state, a count
rate of 5 count s$^{-1}$ (0.9 count s$^{-1}$ after pileup and a pileup
fraction $\sim$70\%) is expected.

We first extracted light curves of the NGC 4151 nucleus in each
observations from the PSF core. Data were extracted from a $1.^{\prime
  \prime}5$~radius circular region centered on the source; this is
approximately the 90\% encircled energy radius of the point spread
function (PSF) at 1.49 keV (Figure~\ref{image}).  The (negligible)
background data were taken from a source free, $10^{\prime
  \prime}$~radius circular region of the same chip.  In the same PIMMS
energy band, the observed count rate is between 0.88 and 0.97 count
s$^{-1}$ without pileup correction, implying a significant pileup
fraction.

\begin{figure}[H]
\epsscale{1.2}
\plottwo{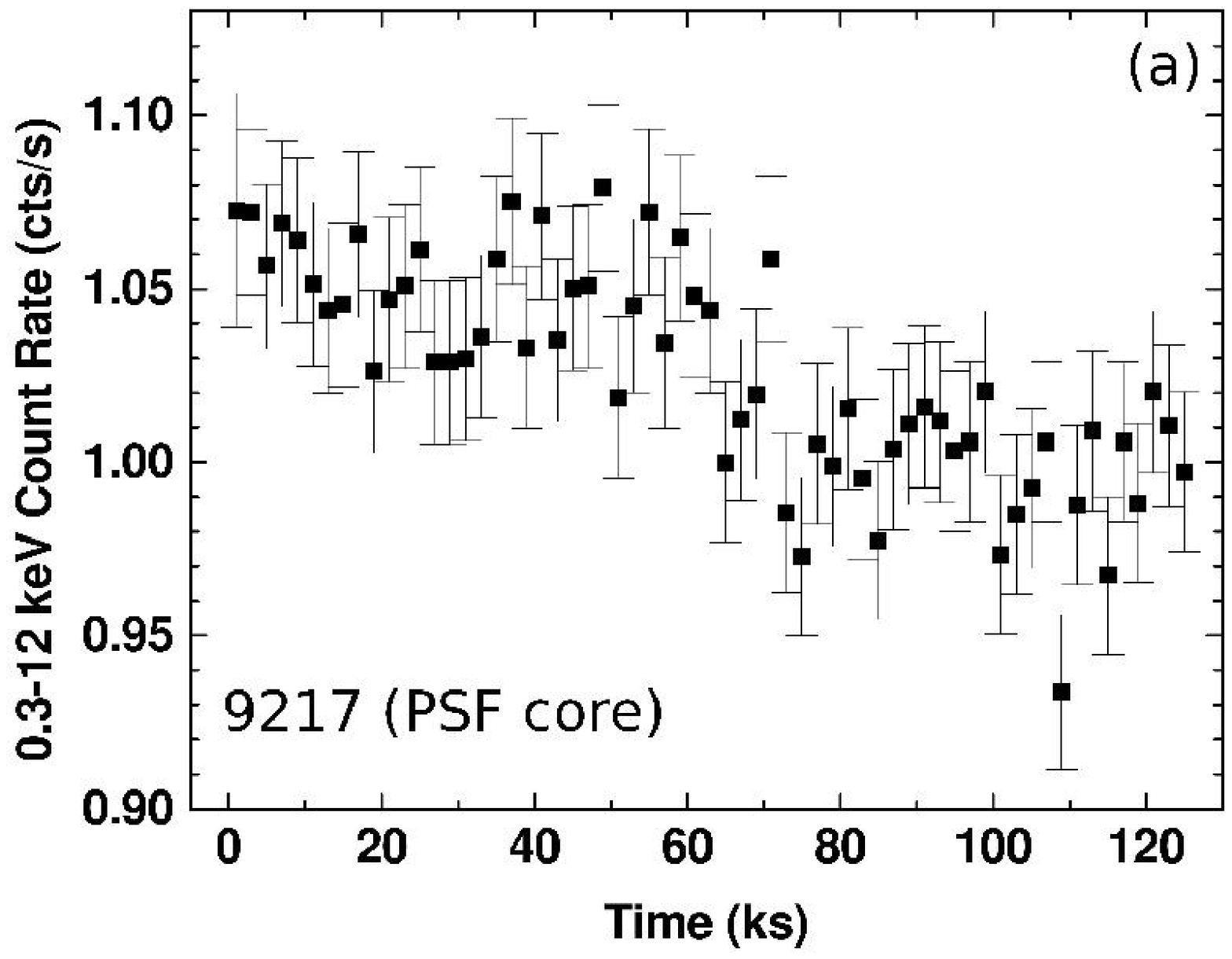}{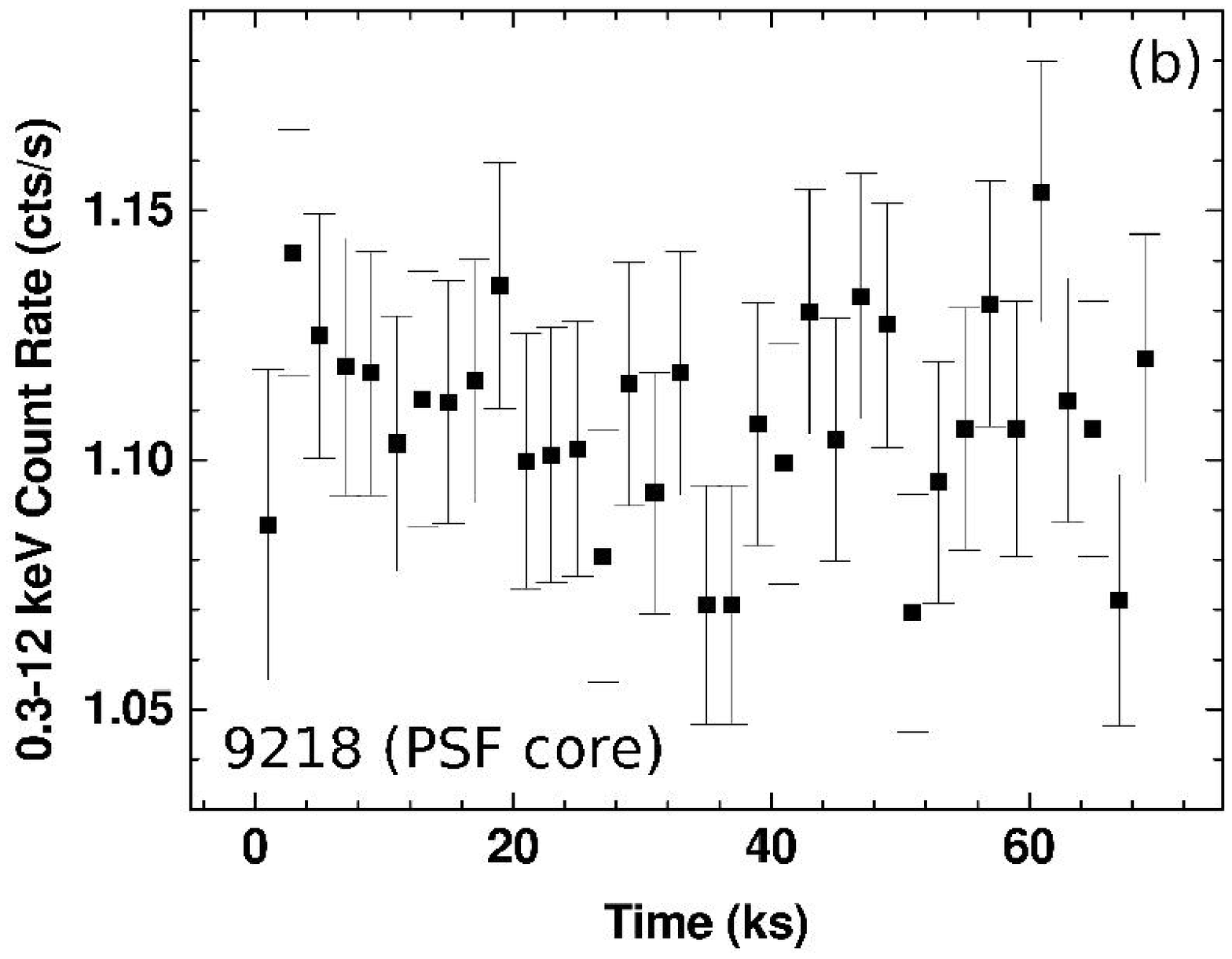}
\plottwo{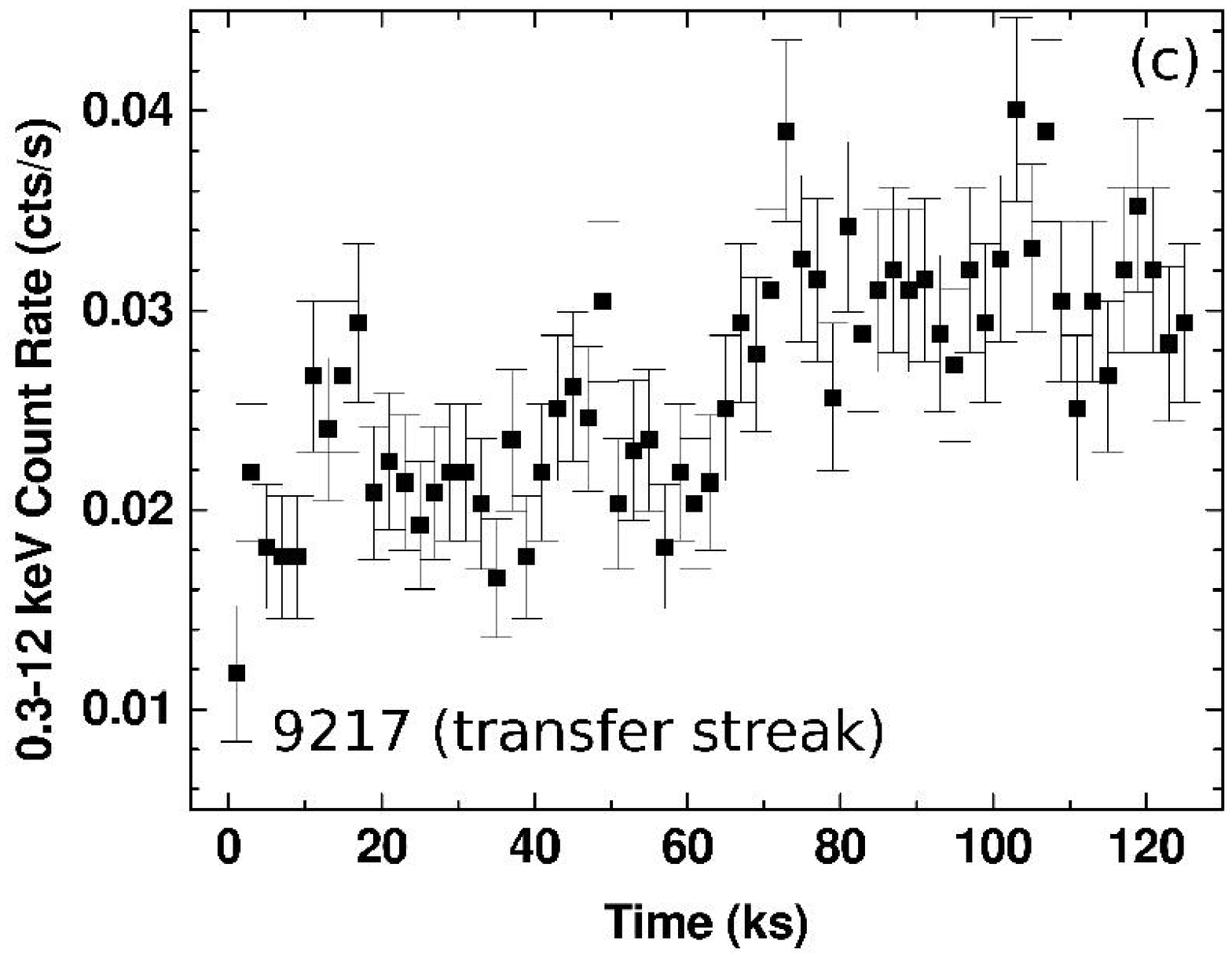}{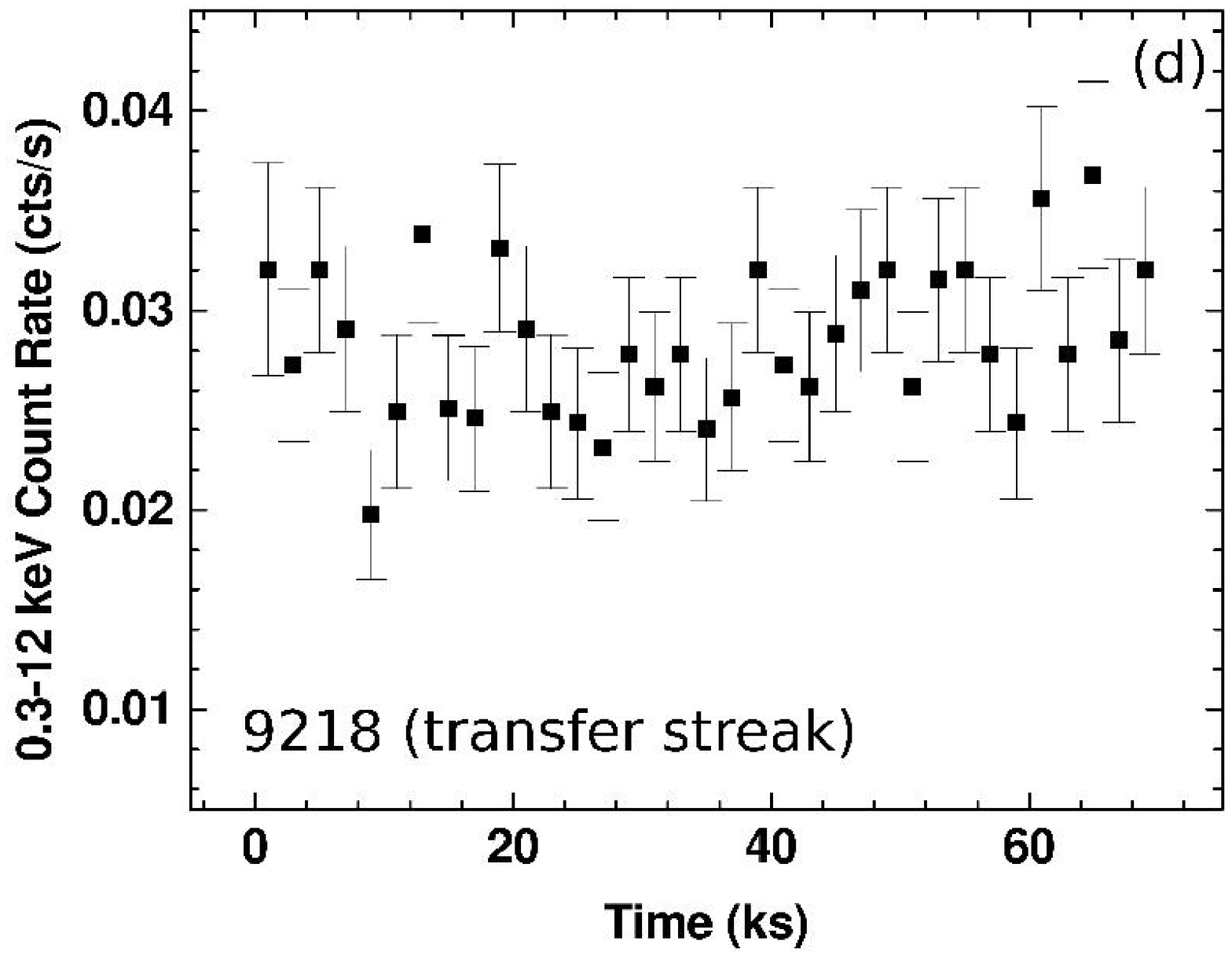}
\plottwo{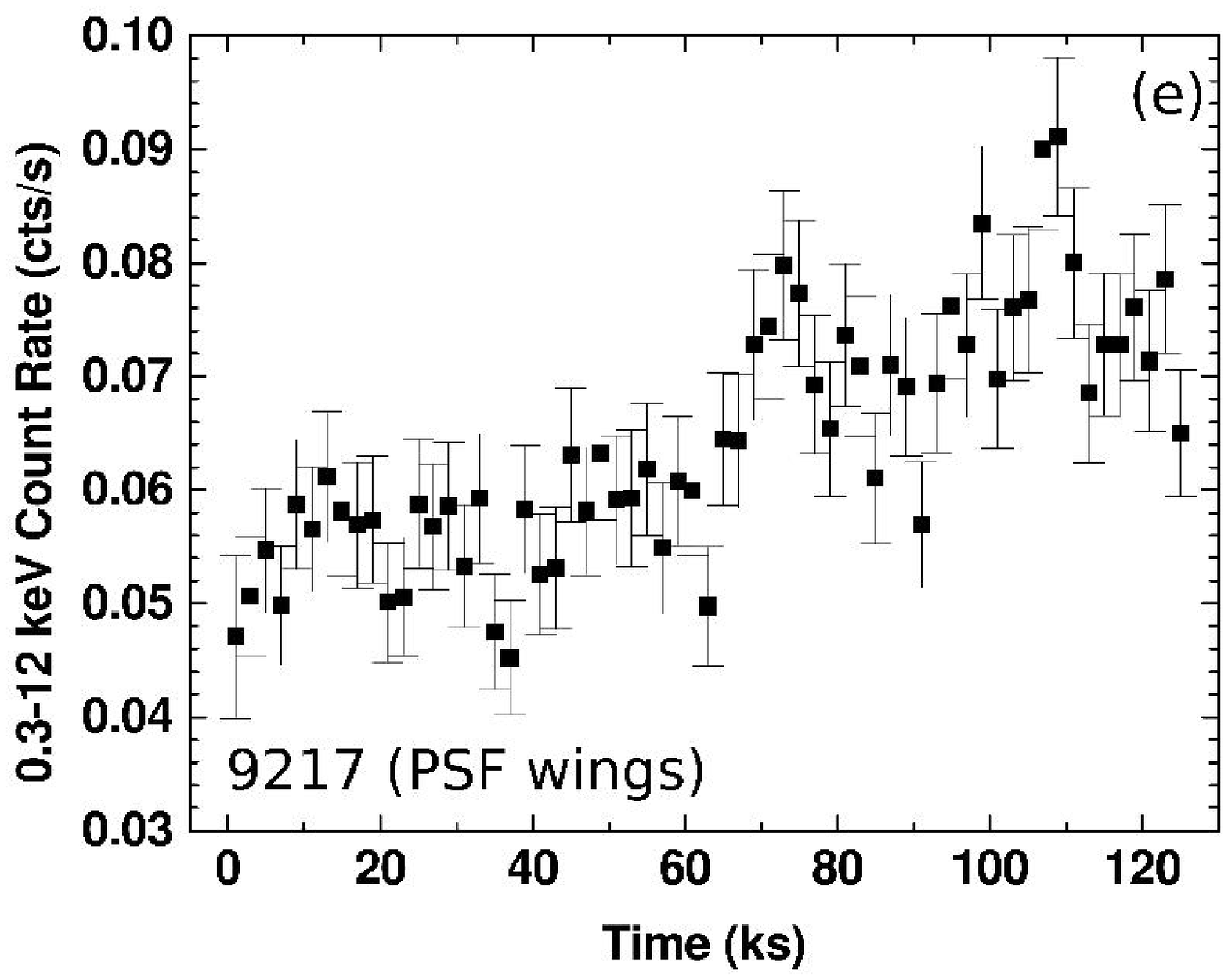}{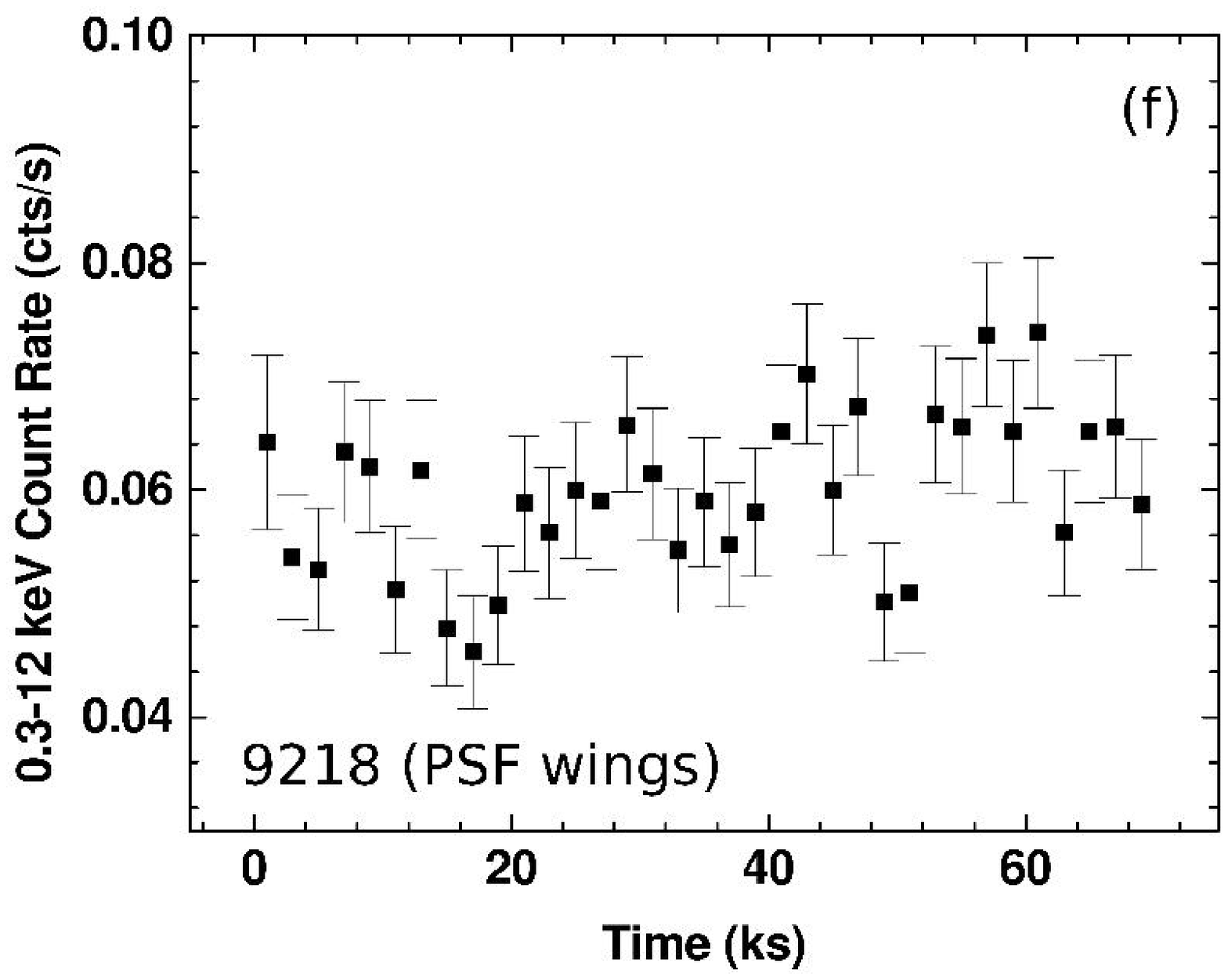}
\caption{NGC 4151 0.3--12 keV light curve for: (a) ObsID 9217 PSF
  core; (b) ObsID 9218 PSF core; (c) ObsID 9217 transfer streak; (d)
  ObsID 9218 transfer streak; (e) ObsID 9217 PSF wings; (f) ObsID 9218
  PSF wings.  Note 0.3--12 keV count rate is used to include high
  energy photons because of pileup.
\label{lc}}
\end{figure}

Figure~\ref{lc} shows the resulting light curves (0.3--12 keV) of the
nucleus for both ObsIDs.  There is a $\sim$5\% decrease in count rate
over $\sim$10 ks during the second part of ObsID 9217.  At first
glance it may reflect a real decrease of the intrinsic source flux,
due to enhanced obscuration of the nucleus as seen in some other
Seyfert galaxies (e.g., Mrk 766, Turner et al. 2006; NGC 1365,
Risaliti et al. 2009a,b).  However, it is known that the relationship
between the source flux and the ACIS count rate is non-linear for a
bright source due to heavy pileup.  An increased source flux may
result in a reduced count rate, causing more severe pileup (sometimes
a hole in the center of a bright source).  We need to be aware of this
possibility, and the following checking is applied.

The events from CCD readout streaks are not affected by the pile up,
because of the very short ``frame time'' (40$\mu$s to clock out
electrons from one row to another) and the spreading of photons over a
larger area.  Using CIAO tool {\it acisreadcorr}\footnote{\url{See
    http://cxc.harvard.edu/ciao/threads/acisreadcorr/index.html}.
  With subarray observations, the readers are reminded that there is a
  known bug in the current version of {\it acisreadcorr} requiring
  modify the {\tt BACKSCAL} header keyword.}, the out-of-time source
events can be identified with a {\tt STATUS} value.  Light curves were
extracted for these unpiled photons in the ACIS readout streak and
shown in Figure~\ref{lc}.  There is a 50\% increase in source count
rate $\sim 60$ks after the starting of ObsID 9217, at the time of the
apparent decrease in the PSF core count rate.  To compute the source
count rate from the readout streak, we used the formula in Marshall et
al. (2005) for the accumulated exposure of a readout streak ($t_s$),
giving a $t_s=990$ s for ObsID 9217 and $t_s=487$ s for ObsID 9218.
The inferred source count rate is 2.3 cps and 3.5 cps for the first
and second part of ObsID 9217, respectively, and 3.2 cps during ObsID
9218.

A second check is offered by the wing of PSF which contains very
little encircled energy and hence a low count rate spread over a much
larger area.  Therefore the events in the PSF wings are not affected
by pileup.  We extracted light curves from a ring region centered on
the nucleus (with an inner radius of 5 pixel and an outer radius of 9
pixel; Figure~\ref{image}) in the outer nucleus region, which are
shown in Figure~\ref{lc}.  Based on the encircled energy fraction, we
expect the extracted count rate of the outer PSF region represents
approximately 2\% of the total count rate level (see
\S~\ref{psfcorr}).  The source count rate inferred from this method is
2.8 cps and 3.6 cps for the first and second part of ObsID 9217,
respectively, and 3.0 cps during ObsID 9218, which indicates that the
observed lower count rate in the first part of ObsID 9217 is due to an
intrinsic low incident flux rather than the response of pileup to a
higher flux.

The readout streak and PSF wings count rates agree well.  The
discrepancy between the observed count rates and these inferred true
count rates indicates a significant pileup.  PIMMS simulation
indicates that the PSF core suffers from $\sim$30--50\% pileup, which
is consistent with the pileup fraction estimated from the spectral
fitting in the following section.

\subsection{Diagnostic Analysis with Empirical Spectral Modeling}\label{specdiag}

A pileup mitigation method, described in great details in Davis
(2001), is widely used and has been implemented in the spectral
modeling tools such as {\em XSPEC} (Arnaud 1996), {\em Sherpa}
(Freeman et al. 2001).  The effectiveness of the model is demonstrated
by recovering identical spectral parameters from the moderately piled
zeroth-order data of quasar S5 0836+7104 to those derived from unpiled
HETG grating spectrum (Davis 2001).  Next we use the Davis (2001)
pileup model to derive spectral parameters from fitting the piled PSF
core spectra and compare with unpiled spectra extracted from PSF wings
and transfer streak.  To apply the pileup model, it is important to
analyze spectra extracted from observations with nearly constant count
rate.  We therefore divided ObsID 9217 into two segments separated at $t=57$ ks
based on the light curve (Figure~\ref{lc}), dubbed 9217a (high PSF
core count rate) and 9217b (low PSF core count rate).

Ogle et al. (2000) and Yang et al.(2001) both found that a hard power
law absorbed by a large column and a soft power law absorbed by the
Galactic column ($\sim 2\times 10^{20}$ cm$^{-2}$, Murphy et al. 1996)
are needed to describe the continuum of NGC 4151 nucleus.  We adopt
the same dual power law model in the spectral fitting (hereafter,
$\Gamma_1$ and $\Gamma_2$ is the photon index for the soft and the
hard absorbed power law component, respectively).  Following Yang et
al. (2001), narrow gaussian features are also added to fit the most
prominent emission lines in the spectra.  An emission line feature at
6.37 keV is identified with a FeK$\alpha$ line from neutral or
weakly-ionized material.

{\bf A. Modeling the PSF Core Spectra with Pileup --} Source and
background Pulse Invariant (PI) spectra and the associated Auxiliary
Response File (ARF) and Redistribution Matrix File (RMF) files were
generated with CIAO tool {\em specextract}.  A circular extraction
region with a radius of 3 ACIS pixels ($1.5\arcsec$) was used, which
corresponds to a PSF fraction of 90\% (1.49 keV). The spectra were
grouped to a minimum of 25 counts per bin to use $\chi^2$ statistics.
Fitting statistics were evaluated including channels above 0.3 keV, up
to 11 keV.  Davis (2008, 6th CIAO workshop) emphasized that, it is
important to include the available photons in the hardest energy
range, even where {\em Chandra}'s HRMA has little effective area.  As
these photons can only come from piled low energy photons, they are
particularly useful to constrain the pileup model.

\begin{figure}[H]
\centerline{\includegraphics[angle=0,width=0.5\textwidth,height=0.4\textwidth]{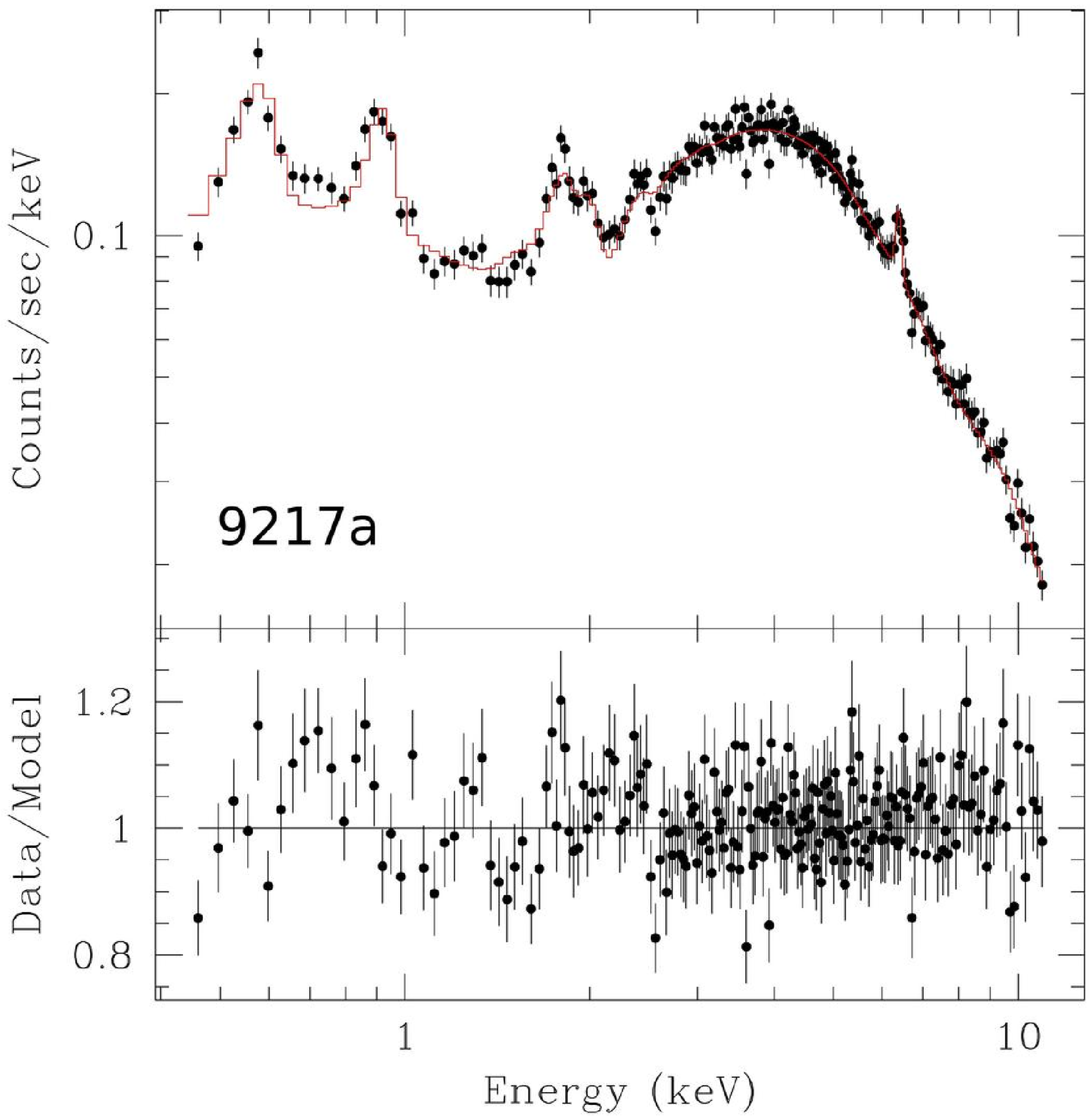}}
\centerline{\includegraphics[angle=0,width=0.5\textwidth,height=0.4\textwidth]{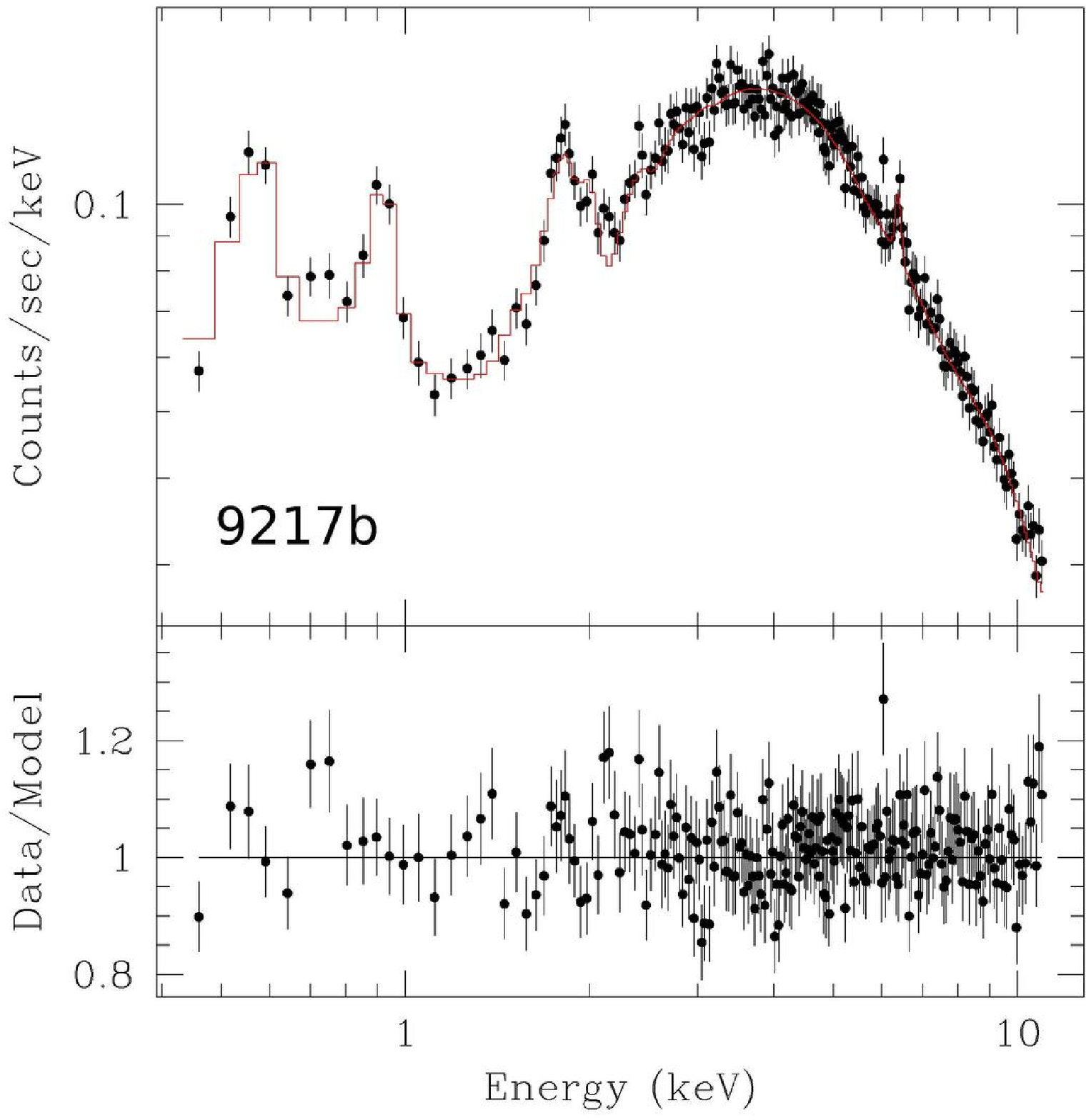}}
\centerline{\includegraphics[angle=0,width=0.5\textwidth,height=0.4\textwidth]{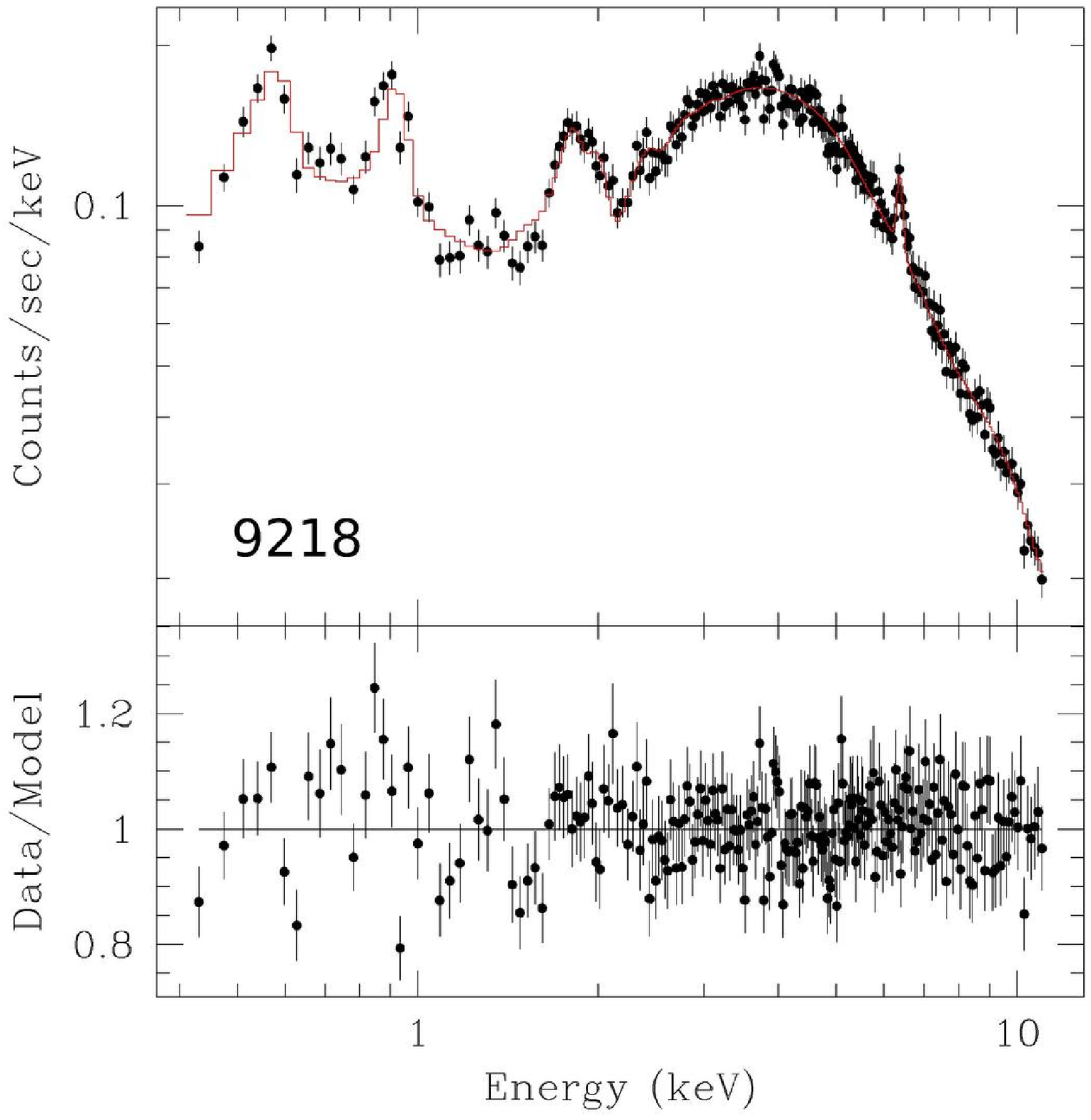}}
\caption{The PSF core spectra and spectral fits using pileup model
  with two absorbed power law components. \label{simple}}
\end{figure}

If we blindly fit the apparently piled spectra with standard model
such as a simple two power-law components, we get an extremely flat
photon index $\Gamma_2\sim -0.1$ for the hard component ($\Gamma_1
\sim 2.5$ for the soft component).  A similar hard photon index was
derived in the piled PSF core spectra in Yang et al. (2001).  When the
pileup model ({\em jdpileup}, Davis 2001) is applied, the corrected,
still rather flat spectral indices are $\Gamma_2$=$0.68\pm 0.05$ (90\%
confidence interval for one interesting parameter), $0.93\pm 0.08$,
and $0.89\pm 0.07$ for ObsIDs 9217a, 9217b, and 9218, respectively.
The resulting fits are shown in Figure~\ref{simple}.

{\bf B. Modeling the PSF Wings Spectra --}\label{psfcorr}
Alternatively, we can extract unpiled photons from outer region at
larger radial distance from the bright nucleus.  We adopted the same
extraction annulus that was used to extract light curve of the PSF
wings (see \S~\ref{ltcv}).  Because of the luminous nucleus, the
extracted spectrum still have enough photons to determine spectral
slope.

\begin{figure}[H]
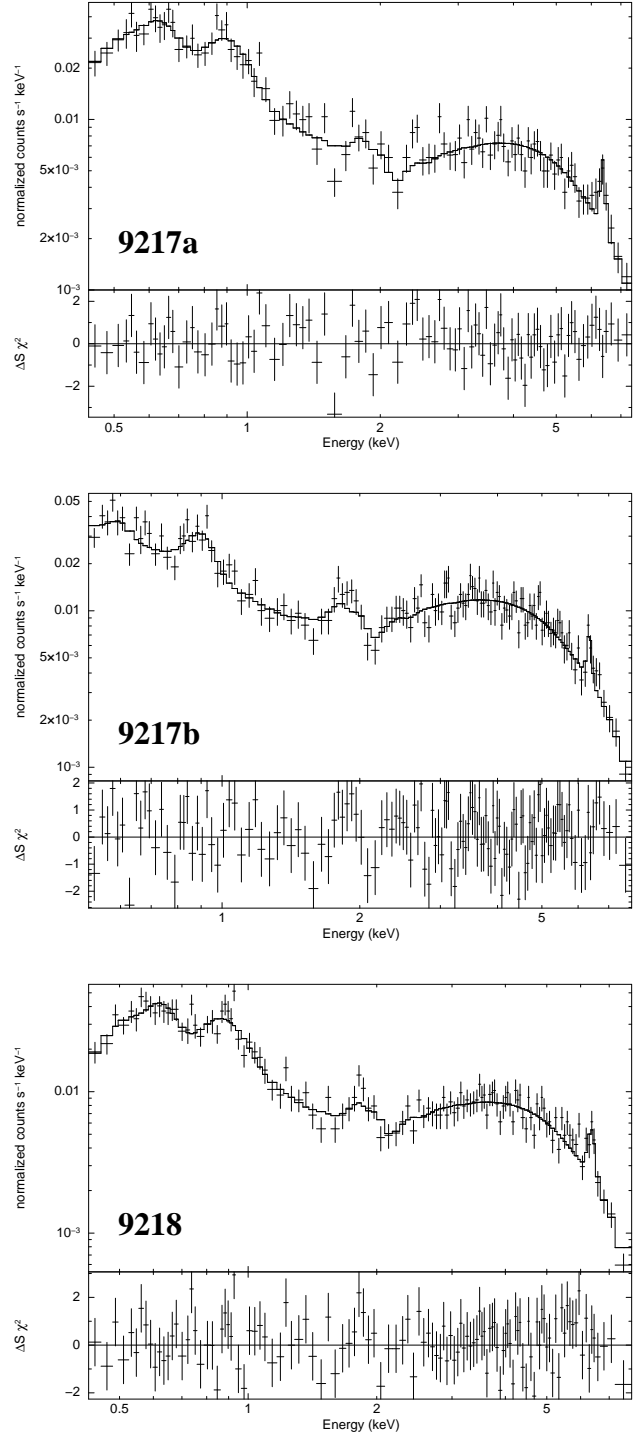

\centerline{\includegraphics[angle=-90,width=0.5\textwidth]{f5a.eps}}
\centerline{\includegraphics[angle=-90,width=0.5\textwidth]{f5b.eps}}
\centerline{\includegraphics[angle=-90,width=0.5\textwidth]{f5c.eps}}
\caption{Spectral fits using the unpiled PSF wings for the three segments. \label{psffit}}
\end{figure}

To model the spectra extracted from events only in the outskirts of
the PSF, we need to correct the ARF for the extraction region of a
partial PSF coverage.  This correction is energy dependent and can be
done through simulation using {\em Chandra} HRMA ray-tracing simulator
(ChaRT\footnote{\url{http://cxc.harvard.edu/chart/}}) and
MARX\footnote{\url{http://space.mit.edu/ASC/MARX/}}.  The detector
position (DETX, DETY) of the nucleus was converted to the set of
$\theta$ (off-axis angle) and $\Phi$ (rotation angle) and input to
ChaRT, then the ray-tracing was used with our observation settings to
create an event file with MARX.  The simulated events from the annulus
region were extracted, and a correction function can be
derived\footnote{Details including verification of this procedure are
  available at
  \url{http://www.astro.isas.jaxa.jp/$\sim$tsujimot/arfcorr.html}}.  We
find photon index $\Gamma_2$=$0.43^{+0.28}_{-0.37}$,
$0.47^{+0.18}_{-0.09}$, and $0.69^{+0.28}_{-0.26}$ for ObsIDs 9217a,
9217b, and 9218, respectively. The resulting fits with absorbed power
law models and emission lines are shown in Figure~\ref{psffit}.

We caution that the extended emission contributes to the extracted
annulus when modeling the PSF wings spectra.  To quantify this
contamination, first, we made use of the PSF deconvolved HRC data
(Wang et al. 2009) and compared the extended counts in the extracted
annulus to the nucleus.  Second, we took the readout streak spectra as
the nuclear spectra, and simulated the nucleus as a point source with
ChaRT and MARX.  The extracted counts from the simulated events (PSF
wings) were compared with the observed wing region (PSF wings and the
extended soft emission).  Both indicate the contamination from
extended emission is $\sim$10\% of the PSF wings and insignificant.

\begin{figure}[H]
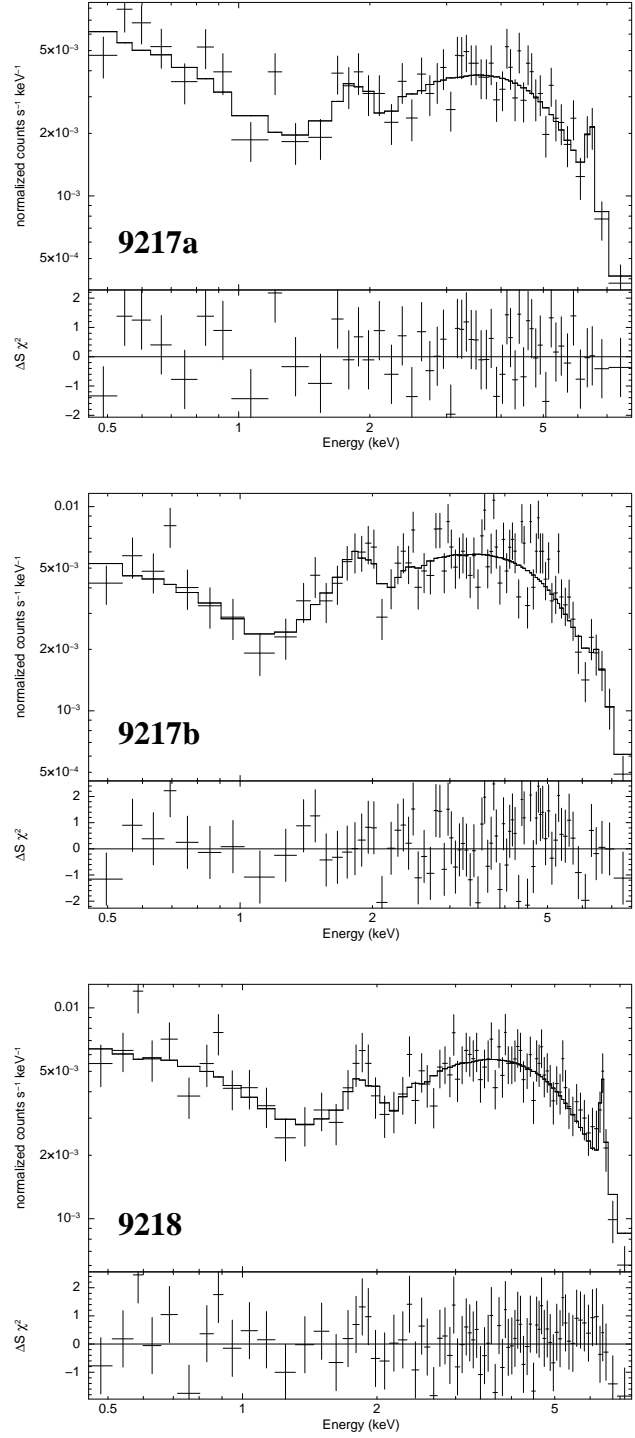

\centerline{\includegraphics[angle=-90,width=0.5\textwidth]{f6a.eps}}
\centerline{\includegraphics[angle=-90,width=0.5\textwidth]{f6b.eps}}
\centerline{\includegraphics[angle=-90,width=0.5\textwidth]{f6c.eps}}
\caption{Spectral fits using the unpiled ACIS readout streak for the three segments. \label{streak}}
\end{figure}

{\bf C. Modeling the Transfer Streak Spectra --} To verify the results
from the pileup model, we extracted source spectrum from the ACIS
readout streak. Following the prescription of Smith et al. (2002), in
calculating the effective area and the detector response we assumed
that the events in the transfer streak originated at the on-axis
source position.  Similar approaches can be found in analysis of {\em
  Chandra}/ACIS observations of NGC 6251 (Evans et al. 2005) and NGC
2992 (Colbert et al. 2005).  The X-ray spectra were extracted using
photons flagged as source events by {\it acisreadcorr}.  The photon
index from the best-fit model is $\Gamma_2$=$0.74^{+0.09}_{-0.05}$,
$0.86^{+0.06}_{-0.10}$ and $0.77\pm 0.13$ for ObsIDs 9217a, 9217b and
9218, respectively. The resulting fits with simple absorbed power law
models and emission lines are shown in Figure~\ref{streak}.

The fitting parameters of interest for the source spectra extracted
using the three different methods are summarized in
Table~\ref{simple_tab}.  Both indirect measurements of the nucleus
spectra are consistent with the results of fitting the PSF core
directly with the pileup model (e.g., for ObsID 9218,
$\Gamma_{2,core}=0.89\pm0.07$, $\Gamma_{2,wing}=0.69^{+0.28}_{-0.26}$,
$\Gamma_{2,streak}=0.77\pm0.13$).  Even flatter spectral states have
also been reported in previous unpiled {\em Chandra} observations
($\Gamma_{2,core}=0.4\pm0.3$, Ogle et al. 2000;
$\Gamma_{2,core}=0.32^{+0.05}_{-0.12}$, Yang et al. 2001), derived
using the same spectral model.  This gives us confidence to attempt
fitting the PSF core with more physically meaningful components while
applying the pileup correction, as described in the next section.

Table~\ref{simple_tab} also supports that there is consistent
variations in the intrinsic flux among the three ObsIDs, independent
of which spectral fitting method is used.  As indicated by the unpiled
count rates in \S~\ref{ltcv}, the flux in ObsID 9217a is lower
compared to that in ObsID 9217b.

\section{Spectral Modeling with Reflection and Absorption}\label{models}

The most notable result from the spectral analysis above is the flat
power law that characterizes the 2--10 keV spectrum of the NGC 4151
nucleus.  As the circumnuclear region of an AGN is a complex mixture
of emitting and absorbing materials (see Turner \& Miller 2009 for a
review), parameters from the satisfactory fitting with the simple
phenomenological model adopted in \S~\ref{specdiag} does not
necessarily represent the intrinsic spectral properties of the nucleus
and instead should be used with caution to further interpret the
spectral variability.

The variable X-ray spectra of AGN (low-flux, hard-spectrum state and
high-flux, soft-spectrum state) have been interpreted with more
complex models than simple power-law (e.g., Schurch \& Warwick 2002;
Miller et al. 2007).  Reprocessing of some of the X-ray continuum by
optically thick material in the accretion disk can create the
``Compton reflection hump''.  If the intrinsic flux of the continuum
decreases and the hard reflection component remains constant (e.g.,
there is a long delay before the reflection component responds to the
lower flux due to the light travel time), a low flux state with a
hardened spectrum of the nucleus can be observed.

Almost indistinguishable from the reflection model, the same low flux,
hard spectrum state can be observed if a dense cloud partially
obscures the continuum source.  Varying covering factor of partial
covering cloud(s) in the line of sight can thus also account for the
spectral variability (e.g., Risaliti et al. 2009a).  Next, we try to
develop more physical description of the {\em Chandra} spectra with
reflection and partial covering components.

\subsection{Modeling the Reflection Component}\label{ref_model}

\begin{figure}[H]
\centerline{\includegraphics[angle=-90,width=1\textwidth]{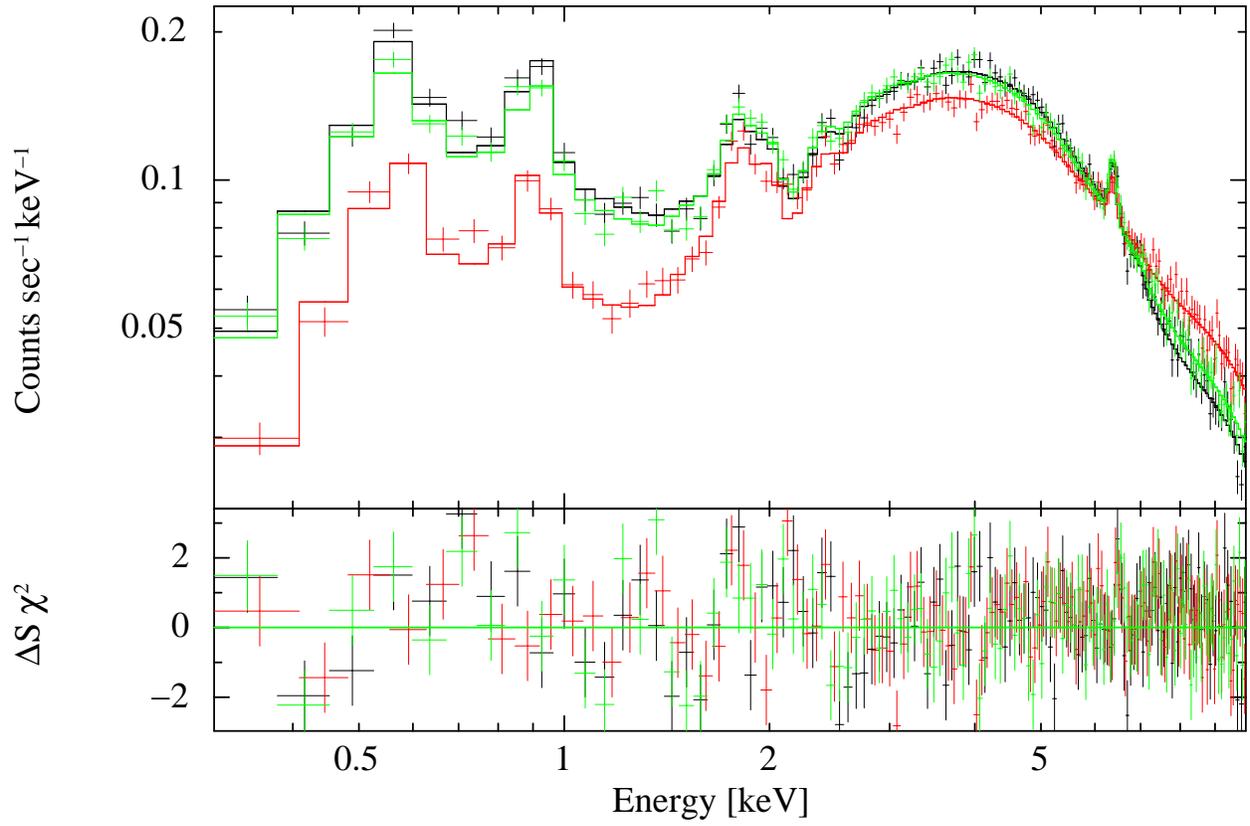}}
\caption{The spectral model with a Compton reflection component, derived by fitting the three segments simultaneously. See Table~\ref{reflect_tab} for fitting parameters. \label{reflect}}
\end{figure}

The PSF core spectra were fitted with an absorbed power law continuum
and an additional steep power law to model the soft excess.  A Compton
reflection component is added with the PEXRAV model (Magdziarz \&
Zdziarski 1995), with parameter {\tt rel\_refl}, the reflection
scaling factor, fixed at $-1$.  The hard power law component and the
reflection component have the same spectral slope.  The same narrow
gaussian components as in previous empirical fitting
(\S~\ref{specdiag}) were added to improve the fit of the emission line
features.  All spectra in the three segments of observations are
fitted simultaneously.  Since the Compton reflection component is not
expected to vary on the time scale of our observation, the photon
indices of the reflection component in the three segments are linked
and left free to vary.  The ratio $R$ between the normalization of
reflection component and the normalization of continuum component
allows us to evaluate the strength of reflection.  The spectral
fitting parameters of interest are summarized in
Table~\ref{reflect_tab}, and the resulting fits for the three
observation segments are shown in Figure~\ref{reflect}.  The best-fit
photon index of the hard continuum is $\Gamma_2=1.15^{+0.08}_{-0.17}$
with this model ($\chi^2$/d.o.f=2260/2033).

Since most of the nuclear soft emission seen by ACIS originates from
blended unresolved lines (see the {\em Chandra}/HETG spectrum, Ogle et
al. 2000), alternatively we attempted another model with the HETG
emission line data frozen to the measurements in Ogle et al. (2000)
plus varying continuum components.  This may be more accurate for the
soft emission affected by the pileup, assuming that the emission line
spectrum is unchanging.  The spectral fitting parameters of interest
are also summarized in Table~\ref{reflect_tab}, and the
$\Gamma_2=1.09\pm 0.09$ is very close to the above simple model. The
resulting goodness of fit $\chi^2$/d.o.f is 2380/2033.

%We fix the photon index of the power-law at 1.7.

\subsection{Effect of Partial Covering}\label{partial}

Complex variable X-ray absorption associated with NGC 4151 has long
been identified (e.g., Ives et al. 1976, Weaver et al. 1994a,b,
Schurch \& Warwick 2002, Kraemer et al. 2005, Puccetti et
al. 2007). Partial covering of the continuum source obscures a
fraction of the soft continuum and hence gives the appearance of a
much harder spectrum. Building on the above model, we add a partial
absorber
(PCFABS\footnote{\url{http://heasarc.gsfc.nasa.gov/xanadu/xspec/manual/XSmodelPcfabs.html}})
to the uniformly absorbed hard power law continuum component.  Again
all three segments are fitted simultaneously.  The absorbing column
densities of the partial obscuring cloud are linked and free to vary.
This constrains the variation of the covering factor.  The spectral
fitting results with the partial covering component are summarized in
Table~\ref{pcfabs}.  The model reproduces the spectra well, with the
best-fit $\Gamma_2=1.68^{+0.05}_{-0.15}$ close to the ``canonical''
value for Seyfert 1 AGNs ($\chi^2_\nu$=2219/2029).  The covering
factor varied between 37\% and 49\% during our observation, causing a
decreased flux in ObsID 9217a.

Using the alternative model with frozen HETG lines (\S~\ref{ref_model}) does not significantly change the fitting results.  The spectral fitting parameters of interest are also summarized in Table~\ref{reflect_tab}, and the resulting goodness of fit $\chi^2$/d.o.f is 2372/2029. The $\Gamma_2$ is very close to the above simple model. There are similar variations in the covering factors.  

\subsection{Modeling an Ionized Absorber}\label{warm_model}

\begin{figure}[H]
\centerline{\includegraphics[angle=-90,width=1\textwidth]{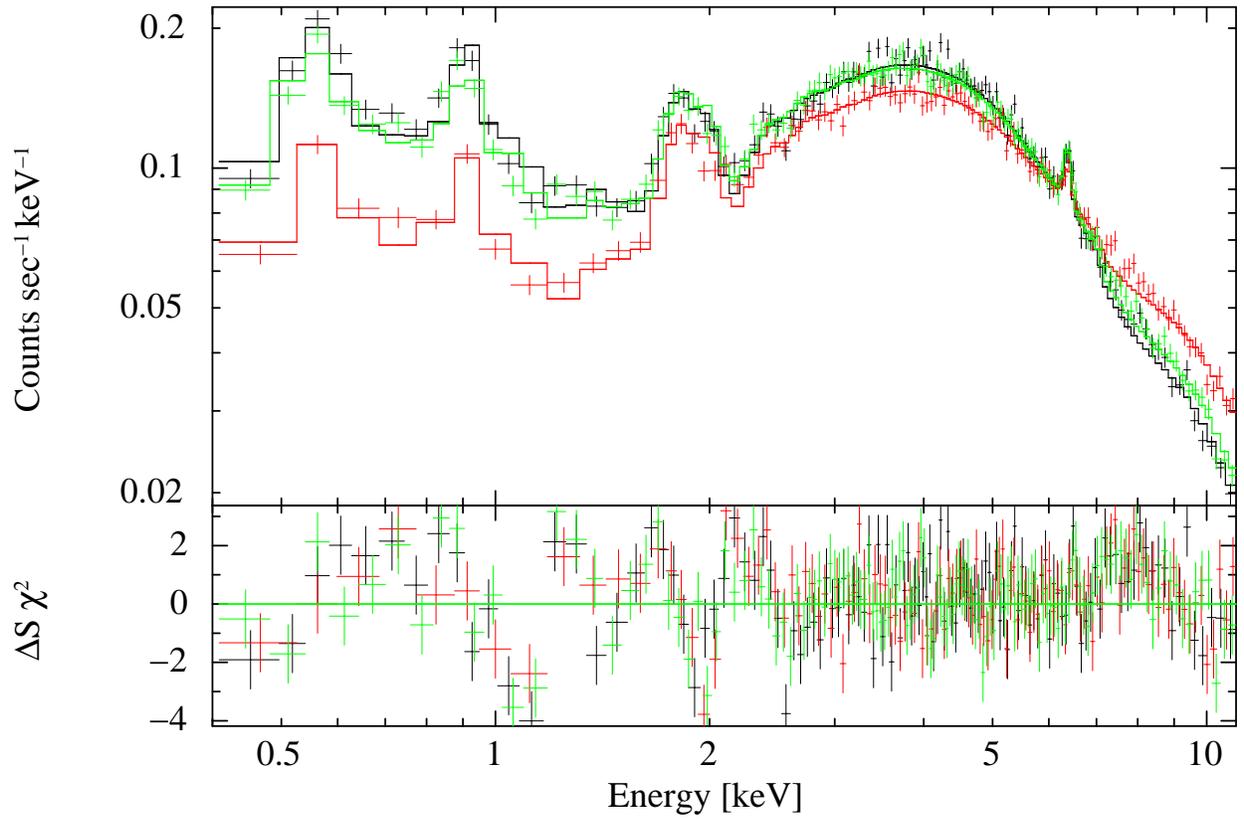}}
\caption{The spectral model with ionized absorber and a Compton reflection component, derived by fitting the three segments simultaneously.  See Table~\ref{absori_tab} for fitting parameters.\label{absori_fig}}
\end{figure}

Given that mostly the low-energy part of the spectrum is subjected to
the ``warm absorber'' and that the current data are considerably
affected by pileup, we did not attempt fitting with complex
photoionization models to the CCD resolution ACIS spectra.  We have
replaced the partial covering absorber in \S~\ref{partial} with an
XSPEC model component
ABSORI\footnote{\url{http://heasarc.gsfc.nasa.gov/docs/xanadu/xspec/manual/XSmodelAbsori.html}}
(Done et al. 1992, Zdziarski et al. 1995) assuming absorption from a
simple ionized absorber present in the line of sight to the active
nucleus.  Again the soft emission is modeled with a power law
continuum with fixed HETG lines.  The modeling is further simplified
by freezing the photon index of the hard component at $\Gamma=1.65$
(Schurch \& Warwick 2002).  The resulting fit is shown in
Figure~\ref{absori_fig}, and the fitting parameters are summarized in
Table~\ref{absori_tab} (goodness of fit $\chi^2/d.o.f=2429/2031$).
The ionization parameter of the warm gas, $\log \xi=0.8-1.1$, is in
agreement with the results in De Rosa et al.\ (2007) and Armentrout et
al.\ (2007).

\section{Revisiting the {\em ASCA} Long-Look}\label{asca_look}

It is interesting to note that we have again observed NGC 4151 in a relatively
low state ($F_{2-10keV}\sim 6\times 10^{-11}$ erg s$^{-1}$ cm$^{-2}$),
similar to two earlier {\em Chandra} Observations (Ogle et al. 2000,
Yang et al. 2001).  {\em XMM}-Newton also observed NGC 4151 three times
December 21-23, 2000.  The absorption corrected 2--10 keV flux was
$5.8\times 10^{-11}$ erg s$^{-1}$ cm$^{-2}$ (Schurch et al. 2003),
indicating the nucleus was in a relatively low-flux state similar to
that observed with {\em Chandra} in Ogle et al. (2000) and Yang et
al. (2001).

\begin{figure}[H]
\epsscale{0.6}
\plotone{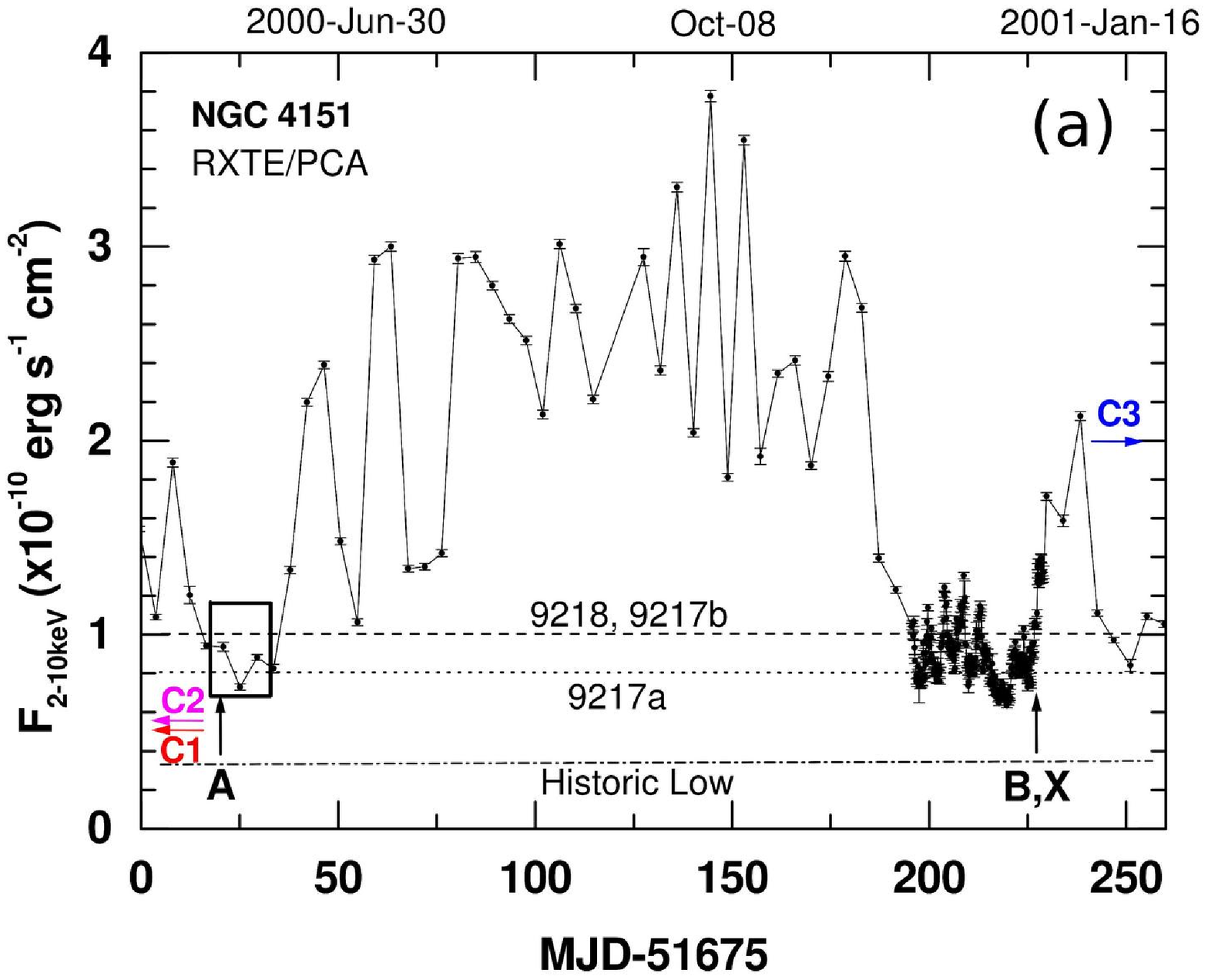}
\plotone{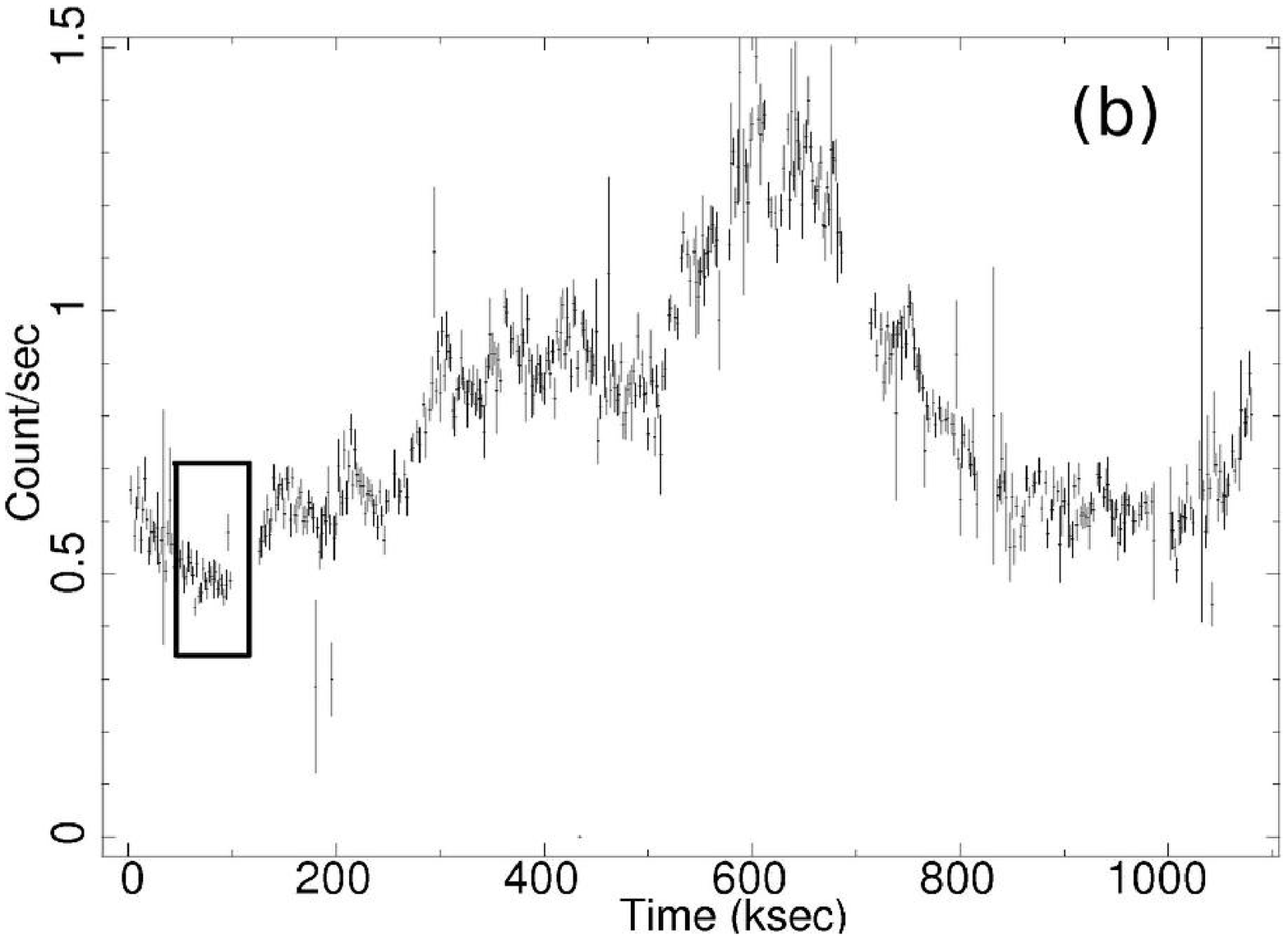}
\caption{(a) The {\em RXTE}/PCA X-ray light curve (2-10 keV) during year
  2000. The time spans when {\em ASCA} long look (2000-May, ``A''; the
  segment in the box is shown in panel $b$), {\em XMM}-Newton (``X''), and
  {\em Beppo}SAX (``B'') observations (2000-Dec) took place happen to cover
  low flux state of NGC 4151.  Earlier {\em Chandra} observations reported
  in Yang et al. (2001), Ogle et al.(2000), and Kraemer et al.(2005)
  are indicated by ``C1'', ``C2'', and ``C3'', respectively. The
  absorption corrected 2--10 keV fluxes for ObsID 9217a (dotted line),
  9217b and 9218 (dashed line) are indicated. The historic minimum
  flux seen with EXOSAT is shown as the dash-dotted line (Pounds et
  al. 1986). (b) The GIS X-ray light curve during the ``long-look''
  {\em ASCA} observation of NGC 4151 in a low state in 2000. The box
  outlines a $\sim 60$ ks interval corresponding to the minimum in the
  light curve, during which we extracted spectrum to study the lowest
  flux state.
\label{pca}}
\end{figure}

We then examined the ``long-look'' observation ($\sim$1 Ms) of NGC
4151 with {\em ASCA} during May 12 to 25, 2000.  Figure~\ref{pca}a shows the
{\em RXTE}/PCA light curve during the monitoring of NGC 4151 in 2000 (see
Markowitz et al. 2003).  The {\em RXTE} light curve shows that the long {\em ASCA}
observation caught the source in an overall low state. In addition,
the {\em ASCA} light curve shows significant variations on time scales of a
few hours. In order to investigate the nature of the lowest states, we
extracted a spectrum from the whole observation, and from a 60~ks
interval corresponding to the minimum in the light curve
(Figure~\ref{pca}b).  The data reduction has been performed following
the standard procedure indicated by the {\em ASCA}
team\footnote{\url{http://heasarc.gsfc.nasa.gov/docs/asca/ahp\_analysis.html}}.
We used screened events provided by the {\em ASCA} archive for the GIS
intruments to extract the light curves and spectra from a circular
region with a $3\arcmin$ radius. The background was extracted from a
source free region in the same field of view. The instrumental
response files were created using the dedicated tools within the {\em
  FTOOLS}
package\footnote{\url{http://heasarc.nasa.gov/docs/software/ftools/}}.

We analyzed both the total and the low-state spectra adopting two
different models for the hard ($>$2~keV) emission: a simple absorbed
power law, and the partial covering+reflection model described in
\S~\ref{partial}. In all cases we obtained a statistically acceptable
result, with $\chi^2_{\nu}\sim 1$ for the first model, and
$\chi^2_{\nu}\sim 0.9$ for the second one.  Given the low value of the reduced
$\chi^2$ in both cases, we cannot estimate whether the partial-covering fit is better than the
simple power law one only based on the global statistics of the two fits.

The analysis of the best fit spectral parameters provides interesting
insghts on the nature of the X-ray emission: for both spectra, the
partial covering model provides acceptable values for the continuum
slope ($\Gamma=1.9\pm0.05$ for the total spectrum, $\Gamma=1.8\pm0.2$
for the low-state spectrum). The intensity of the reflection component
is R=1$\pm$0.5 for the total spectrum, and R=2.5$\pm1$ for the low
state spectrum. This suggests that the reflected component seen in the
low state spectrum is the echo of an on average higher emission,
consistent with the {\em RXTE} and {\em ASCA} light curves.  The
simple power law model provides a ``normal'' value of the continuum
slope for the total spectrum ($\Gamma=1.7\pm0.1$), but an extremely
flat value for the low state spectrum ($\Gamma=1.2\pm0.2$).  This
confirms our interpretation of the {\em Chandra} spectra, and in
particular that (1) harder spectra are associated to lower flux
states, and (2) that the flat observed emission is not due to an
intrinsic change in the continuum shape, but to absorption effects
and/or to the higher relative weight of the reflection component in
low intrinsic flux states.

\section{Discussion}\label{discussion}

The overall picture of the observed source variation is now improved:
counterintuitive to the apparent light curve, we actually witnessed
the source going from a moderate count rate (ObsID 9218) to a low
count rate (ObsID 9217a) and back to a high count rate (ObsID
9217b). As noted in \S~\ref{models}, the interpretation of the
physical origin of this variation is not straightforward and depends
on the spectral modeling.

\subsection{An Intrinsic Flat Power Law Spectrum?}\label{flat}

There have not been any reports for Seyfert galaxies with spectral
indices as flat as $\Gamma\sim 0.7-0.8$, unless they are in the
Compton thick regime.  The earlier {\em Chandra} result of
$\Gamma_2\sim 0.3$ for NGC 4151, a clear-cut Seyfert 1 galaxy was
noted to be ``unusual'' (Yang et al. 2001), but not further discussed by Ogle et
al. (2000) or Yang et al. (2001).  We note that a few cases of low
$\Gamma$ spectra exist in the literature. For example, a
$\Gamma=0.95\pm 0.24$ was measured for a radio loud quasar PKS
2251+113 in the {\em ASCA} quasar sample (Reeves \& Turner 2000).
Strateva et al. (2008) reported that one of the broad double-peaked
Balmer line emitting quasar SDSS J2125-0813, may also have an unusual
flat X-ray spectrum $\Gamma\sim 1$.

If the measured $\Gamma$ with the empirical, uniformly absorbed two power law components model indeed represents
the slope of the intrinsic NGC 4151 continuum in the low state, it may have some implications for the accretion flow onto the central black
hole.  For example, a radiatively inefficient accretion flow (RIAF; for reviews, see Narayan et al. 1998; Quataert 2001) is capable of producing a flat
hard state X-ray spectrum.  Using a low-flux state 2--10 keV luminosity $L_{2-10keV}\sim 3\times
10^{42}$ erg s$^{-1}$, we derive a $L_{bol}\sim 2\times 10^{43}$ erg
s$^{-1}$ for NGC 4151 from the bolometric correction derived in Marconi et
al.\ (2004).  Taking the latest measurement of the black hole mass
$M_{BH}=4.57^{+0.57}_{-0.47}\times 10^7$M$_{\odot}$ for NGC 4151 from
Bentz et al. (2006) yields an Eddington luminosity of $L_{Edd}\sim
6\times 10^{45}$ erg s$^{-1}$, which implies the central engine of NGC
4151 is radiating at a relatively low efficiency,
$L_{bol}/L_{Edd}=3\times 10^{-3}$.  The ratio
indicates that the accretion rate of the NGC 4151 black hole is close
to the critical value $\dot{M}_{crit}\approx 0.01\dot{M}_{Edd}$ (Narayan et al. 1998, Yuan 2007) for a RIAF.

However, a more plausible interpretation is that the extreme flat
slope is the outcome of over-simplified modeling. At least, the value
of $\Gamma$ is known to depend strongly on the treatment of the
complex absorbers (Weaver et al. 1994a, Zdziarski et al. 2002).  We
have verified that the physically meaningful models (reflection model
discussed in \S~\ref{reflection} and partial covering model discussed
in \S~\ref{partc}) also can fit the readout streak spectra well
($\Gamma_2=0.9\pm0.5$, $\chi^2=1.09$ for the reflection model, and
$\Gamma_2=1.7^{+0.5}_{-0.9}$, $\chi^2=1.07$ for the partial covering
model), yielding consistent $\Gamma$ values derived from the piled
spectra but with poor constraints on the fitting parameters due to
much less counts in the readout streak spectra and complexity of the
models.

\subsection{Compton Reflection Component}\label{reflection}

Although the reflection model (\S~\ref{ref_model}) fits the spectrum
well, the problem of the flat spectral slope remains with the derived
$\Gamma_2\sim 1.15$.  The amount of reflection is characterised by
$R=\Delta \Omega/2\pi$, where $R=1$ is often observed for Seyfert 1s,
meaning the illuminated infinite slab extends a solid angle of $2\pi$
to the central source. It is also referred to as the ratio of the
reflected intensity relative to the expected intensity from the
surface subtending $2\pi$ to the continuum source.  The reflection
seems unusally large, $R$=1.5--2.7, while the Fe$K_{\alpha}$ line EW
is relatively small $\sim 100$ eV, inconsistent with the large $EW$ of
the Fe$K_{\alpha}$ fluorescence line ($\geq 1$ keV) expected for
strong reflection (e.g., Guilbert \& Rees 1988, George \& Fabian
1991).

The Fe line flux is also constant within the errors during our
observation, compared to the variable continuum.  This result is
easily explained in the standard cold reflection model where the
narrow FeK$_{\alpha}$ line is produced far from the primary X-ray
source (e.g. from the putative obscuring torus) and so is not expected
to vary rapidly.  As NGC 4151 shows frequent ``flaring'' in the X-ray
continuum light curve, the large $R$ observed here may result from a
response to a previous X-ray flare produced in the inner disk region
before the start of our {\em Chandra} observations. The response to
changes in the X-ray flux by the illuminated gas could even take
longer than the light-crossing time between the primary X-ray source
and the reflecting material (Nicastro et al. 1999, Nayakshin \&
Kazanas 2002).  For example, Liu et al. (2010) have
analyzed SUZAKU monitoring of NGC 5548 and found no strong correlation
between Fe$K_{\alpha}$ flux and the continuum.

In particular, the reflector could be the inner wall of a torus
(Antonucci 1993; Krolik et al. 1994) a few hundred light-days away
from the nucleus (Suganuma et al. 2006).  The inner size of a dust
torus in NGC 4151 was measured as $48^{+2}_{-3}$ days based on the lag
time between the optical and near-infrared light curves (Minezaki et
al. 2004).  The observed narrow Fe$K_{\alpha}$ fluorescent line is
expected, and is unresolved with {\em Chandra}/HETG ($FWHM=1800\pm
200$ km s$^{-1}$, Ogle et al. 2000).  The high signal-to-noise (S/N)
{\em XMM}-Newton/EPIC spectra suggest a broader intrinsic width
($\sigma_{K\alpha}=33^{+6}_{-5}$ eV, $FWHM\sim 3000$ km s$^{-1}$,
Schurch et al. 2003).  Nevertheless, these values are consistent with
a distant reflector.  Therefore at short time scale (hours to days),
correlation between the continuum flux, the reflection continuum and
the iron line emission is unlikely to be found.

\begin{figure}[H]
\epsscale{0.9}
\plotone{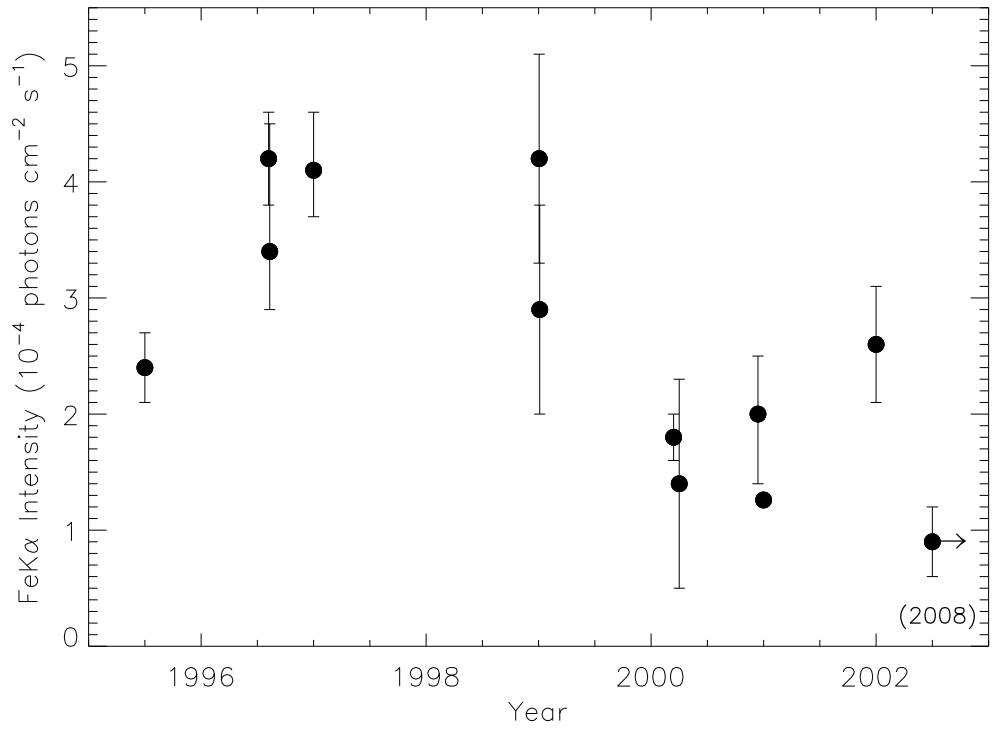}
\caption{Variation of the FeK$\alpha$ line intensity. Data are taken
  from Ogle et al. (2000), Yang et al. (2001), Zdziarski et al.(2002),
  Schurch et al. (2003), and de Rosa et al. (2007).
\label{feline}}
\end{figure}

In Figure~\ref{feline} we show the long-term variation of the
Fe$K_{\alpha}$ line flux based on earlier X-ray observations from {\em
  ASCA}, {\em Beppo}SAX, {\em Chandra}, {\em XMM}-Newton reported in
the literature.  The Fe$K_{\alpha}$ in our observation ($9.7\pm
3.0\times 10^{-5}$ photons cm$^{-2}$ s$^{-1}$) is consistent with that
measured from the {\em XMM}-Newton spectrum ($1.26\pm0.04 \times
10^{-4}$ photons cm$^{-2}$ s$^{-1}$; Schurch et al. 2003) and slightly
lower than the earlier {\em Chandra} results ($1.8\pm0.2 \times
10^{-4}$ photons cm$^{-2}$ s$^{-1}$; Ogle et al. 2000).  Ogle et
al. (2000) attributed nearly 70\% of Fe$K_{\alpha}$ line flux to the
kpc scale extended X-ray emission.  We will examine this claim more
carefully in our study of the ENLR and compare with the lower spatial
resolution $XMM$-Newton study (Schurch et al. 2004).  Significant line
intensity variation is only seen at time scale longer than 1 year,
consistent with an origin from a distant reflector and the result from
{\em RXTE} monitoring (Markowitz et al. 2003).  As suggested in
Markowitz et al. (2003), the variation of Fe$K_{\alpha}$ observed with
{\em RXTE} may still trace the intrinsic continuum variation in the
long term.

Other means of obtaining a larger $R$ are possible, such as: the
light-bending model which takes into account relativistic aberration
(Ballantyne et al. 2003); a peculiar inner disk geometry with deep
funnel; or a partly hidden continuum that is only seen by the
reflector (see Turner \& Miller 2009 for review on strength of the
reflection component).  However, these models apply to X-ray
reprocessing by a Compton-thick inner accretion disk, which is
inconsistent with the fact that neither do we see a broad
Fe$K_{\alpha}$ line, nor a strong Fe$K_{\alpha}$ line with large EW.
It is also inconsistent with the low Eddington ratio of NGC 4151.

Nevertheless, a considerable number of observations of Seyfert 1
galaxies have yielded $R>1$ values (e.g., Mrk 335, Bianchi et
al. 2001, Larsson et al. 2008; Mrk 841, Petrucci et al. 2002).
Notably, Miniutti et al. (2007) obtained $R=2.8\pm 0.9$ for
MCG--6-30-15, which shows the broad iron line and reflection hump
expected from X-ray reflection in the innermost regions of an
accretion disk.  Terashima et al. (2009) found $R\sim 7$ for the
reflection component in NGC 4051 with a modest $EW\sim 140$ eV for FeK
line, similar to the NGC 4151 results.  They suggest that the large
reflection fraction is due to a delayed reflected emission to the
variations between high-flux state and low-flux state.

\subsection{Partial Covering}\label{partc}

To evaluate whether the partial covering component is required by the
{\em Chandra} data, we performed an F-test comparing the statistics
for the reflection model ($\chi^2/dof$ = 2380/2033 = 1.171) and the
partial covering model ($\chi^2/dof$ = 2371/2029 = 1.169) based on the
spectral models with fixed soft emission lines.  The F-test statistic
gives a probability 14\% that suggests only a marginal improvement
over the simpler reflection-only model.  Nevertheless the advantage of
the spectral model consisting of Compton reflection component and an
additional partial covering component is that it provides a spectral
slope $\Gamma_2=1.68$ (\S~\ref{partial}) well within the typical
values of Seyfert 1 AGNs. This suggests that we do not need to invoke
an exotic explanation of the apparent flat spectrum.  The spectral
variations among the three segments could be attributed simply to a
$\sim$10\% change in covering fraction of a high column absorber
towards NGC 4151, although a variation in the ``intrinsic'' continuum
is equally possible (e.g., variation of Compton-thick materials),
given the uncertainties in the fitted values.  Fiore et al. (1990)
argued that the variation of the NGC 4151 spectral slope is intrinsic.
However, De Rosa et al.(2007) found the spectral variability is small
above 10 keV ($\Delta \Gamma=0.2$) using {\em Beppo}SAX data covering
0.1--200 keV.

The presence of a high column density ($N_H\sim 10^{23}-10^{24}$
cm$^{-2}$) X-ray absorber in the line of sight towards the NGC 4151
nucleus is not surprising and observed in previous studies.
Whitehouse \& Cruise (1985) reported a 2-ks dip in the EXOSAT light
curve of NGC 4151, perhaps related to occultation by passage of a star
across the line of sight towards the black hole.  Puccetti et
al. (2007) also concluded that the absorption variability plays a
crucial role in the observed flux variability in NGC 4151.  In another
rare case, Risaliti et al. (2007, 2009b) witnessed occultation by a
Compton-thick cloud crossing the line of sight towards NGC 1365.

We note that the $BeppoSAX$ spectral study of NGC 4151 (De Rosa et
al. 2007) included the same model components as ours, but introduced a
photoionized gas component as warm absorber to the partial covering
cold absorber.  Their results show that, the column density of the
warm absorber is consistent with being constant and only changes over
a time scale longer than months.  It is the covering fraction of the
cold absorber (varying on time scale of days) that dominates the
spectral variability.  Accordingly, the ionization parameter $\xi$
increases when the covering fraction becomes lower, as the warm
absorber exposed to more continuum (see also Schurch \& Warwick 2002).
The presence of the warm absorber will be discussed next.  Our results
(Table~\ref{pcfabs}) quantitatively agree with what De Rosa et
al. (2007) find (e.g., photon index of the hard power law
$\Gamma_2=1.6$, column density of the partial covering absorber
$N_{H,PCFABS}=1.5\times 10^{23}$ cm$^{-2}$, variations in the covering
fraction $f_{cov}$=0.48--0.73 during their observation in 1996).

\subsection{Presence of a Warm Absorber}

The presence of highly ionized gas close to AGN becomes evident in
recent observations by {\em Chandra}, {\em XMM}-Newton, and {\em
  SUZAKU}, where high resolution spectra have shown the presence of
highly ionized absorber as well as lower ionization gas (e.g., Ogle et
al. 2000; Kraemer et al. 2005; Young et al. 2005; Miniutti et
al. 2007; Krongold et al. 2009).

Kraemer et al. (2005) revisited NGC 4151 in 2002 in a relatively low
state (25\% of the historic maximum), again with the $Chandra$ HETG,
focusing on the intrinsic X-ray absorption.  They found that,
comparing to the 2002 data, a combination of higher column density and
lower ionization of the intervening gas may explain the $Chandra$ HETG
spectrum in 2000 with an assumed $\Gamma\sim 1.5$ (see also Armentrout
et al. 2007).  The spectral model consists of five different
absorption components with various radial distances, filtered ionizing
continuum, and covering factor.  It is clearly not feasible to apply
such complex models to our piled spectra.

Our simple ionized absorber fit to the lower-resolution ACIS CCD
spectra suggested an ionization parameter $\log\xi\sim 0.8-1.1$, which
is comparable with the values in De Rosa et al. (2007).  The
variations of the ionization parameters during the three observation
segments hint that the spectral variability may be related to the warm
absorber responding to the varying ionizing luminosity (Schurch \&
Warwick 2002).  We note that the deep absorption edges from ions like
OVII and NeIX will be blended together at low resoltuion and result in
a smooth curvature (Kraemer et al. 2005).  This may explain why our
simpler reflection and partial covering models provide good
description of the ACIS spectra.  Although our reflection and partial
covering models using neutral absorbers are over-simplified given the
presence of warm absorber evident in the HETG spectra, they are
consistent with the ACIS data and not conflicting with the HETG
results.  First of all, the warm absorber model in Kraemer et
al. (2005) focused on reproducing the spectral slope in the hard band
($>$1.5 keV), and excluded fitting the soft energy range because of
the strong emission lines. This leaves room for contribution from a
neutral absorber or a partial covering absorber (e.g., Schurch \&
Warwick 2002, De Rosa et al. 2007).  In a separate {\em XMM}-Newton
RGS grating study of NGC 4151, Armentrout et al. (2007) modeled the
soft emission lines with three photoionized components, with covering
fractions ranging between 0.03--0.70.  Secondly, it is possible that
the amount of low ionization gas can be extremely variable on short
time-scales, as noted in Kraemer et al. (2001).  Although limited by
the signal to noise ratio, studies of short term spectral variability
with high resolution grating spectra will be interesting.

\subsection{Constraints on the X-ray Emission and the Obscuring Clouds}

If the observed X-ray light curve reflects real variation of the
intrinsic continuum of NGC 4151 on time scale of $\sim$10 ks, it is
consistent with the rapid X-ray variability previously reported (Elvis
1976, Lawrence 1980) and also seen in other local Seyfert galaxies
(e.g., Lawrence et al. 1985; Mushotzky et al. 1993; Nandra et
al. 1997).  This variability time scale constrains the X-ray emission
region to be small ($D\simeq 20 r_s$, where $r_s=2GM/c^2$ is
Schwarzschild radius; see Turner \& Miller 2009), assuming
$M_{BH}=4.57^{+0.57}_{-0.47}\times 10^7$M$_{\odot}$ for NGC 4151
(Bentz et al. 2006), thus likely close to the central supermassive
black hole.

If the lower count rate and higher covering fraction during ObsID
9217a is due to motions of clumpy absorbers, then, following Risaliti
et al. (2002), we can constrain the distance of the discrete cloud
from the central black hole assuming that the obscuring clouds are
moving with Keplerian velocities, in terms of the variations in $N_H$.
Adopting $M_{BH}\sim 4.6\times 10^7$M$_{\odot}$, $\rho\sim 10^{10}$
cm$^{-3}$ (typical of broad emission line region [BELR] clouds),
$t\sim 0.7$ day (duration of ObsID 9217a), and $\Delta N_H\sim
10^{23}$ cm$^{-2}$, we derive $r\leq 1.6\times 10^4 r_s$ ($\sim 9$
light-day) from Equation (6) of Risaliti et al. (2002).  This is
consistent with the inferred BELR size ($6\pm 4$ light-day) from
reverberation mapping with H$_{\beta}$ and HeII $\lambda$4686 emission
lines (Peterson \& Cota 1988) and the location of obscuring clouds
derived from {\em Beppo}SAX observation (Puccetti et al. 2007).

\section{Conclusions}\label{summary}

The deep $\sim$200 ks {\em Chandra} observation of the Seyfert 1
nucleus in NGC 4151 was analyzed to understand its complex spectral
variability, in particular the unusally flat photon index reported
from earlier $Chandra$ observations.  We find:

(1) The 2--10 keV flux of the nucleus varied from $6\times 10^{-11}$
erg s$^{-1}$ cm$^{-2}$ (ObsID 9218 and ObsID 9217a) to $\sim 10^{-10}$
erg s$^{-1}$ cm$^{-2}$ (ObsID 9217b), resulting in a non-linear count
rate variation of opposite sign because of significant pileup.

(2) With pileup corrected spectral fitting, we are able to recover the
spectral parameters and find consistency with those derived from
unpiled events in the ACIS readout streak and outer region of the
bright PSF core.

(3) The low flux segment shows a hard photon index $\Gamma_2\sim
0.7-0.9$ similar to that seen in the historical low state for a simple
power-law fit.  More complex, physically meaningful models are
attempted and provide good fits to the piled spectra, including a
Compton reflection model, the reflection model subjected to a
partially covered absorber, and the reflection model subjected to an
ionized absorber.

(4) The observed flat spectrum and its variability can be interpreted
as due to an intrinsically varying continuum with respect to an
underlying Compton reflection component, or a variable X-ray absorber
partially covering the continuum source.  Including the partial
covering absorber provides only a marginal improvement over the
simpler reflection-only model, nevertheless it gives a continuum
photon index that is typical of Seyfert 1s.

(5) If the absorption model is correct, the size of X-ray emission
region is constrained $r\simeq 20 r_s$, and the X-ray absorber is
located at $r\leq 1.6\times 10^4 r_s$ from the nucleus, possibly
associated with the BELR clouds.

(6) If instead the reflection-only model is correct, the presence of a
constant reflection component with respect to the rapidly flaring
continuum of NGC 4151 implies the reflector is located far from the
nucleus, consistent with the absence of broad FeK$\alpha$ fluorescence
line.

\acknowledgements

We thank the anonymous referee for helpful comments that significantly
improved the manuscript.  This work is partially supported from NASA
grant GO8-9101X and NASA Contract NAS8-39073 (CXC). J.W. thanks
Michael McCollough for advice on ACIS readout streaks.  This research
has made use of software provided by the {\em Chandra} X-ray Center
(CXC) in the application packages CIAO and Sherpa.

%\clearpage

\clearpage

\begin{figure}[H]
%\epsscale{0.7}
\centerline{\includegraphics[angle=-90,width=0.8\textwidth]{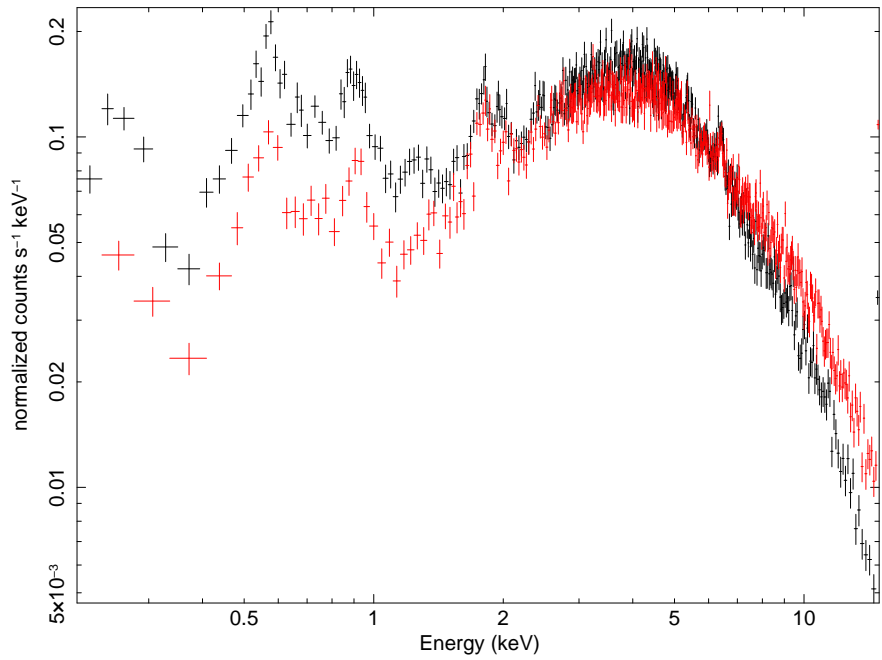}}
\caption{The X-ray spectra of the apparent high count rate segments
  ObsID 9217a (black line) and low count rate segment 9217b (red
  line).  Because of pileup, note how both spectra still have flux
  extending to above 8 keV, the enery range in which {\em Chandra}'s
  effective area drops significantly. The 9217b spectrum shows a
  harder tail, suggesting more severe pileup than 9217a.
\label{twospec}}
\end{figure}

\begin{figure}[H]
\centerline{\includegraphics[angle=-90,width=1\textwidth]{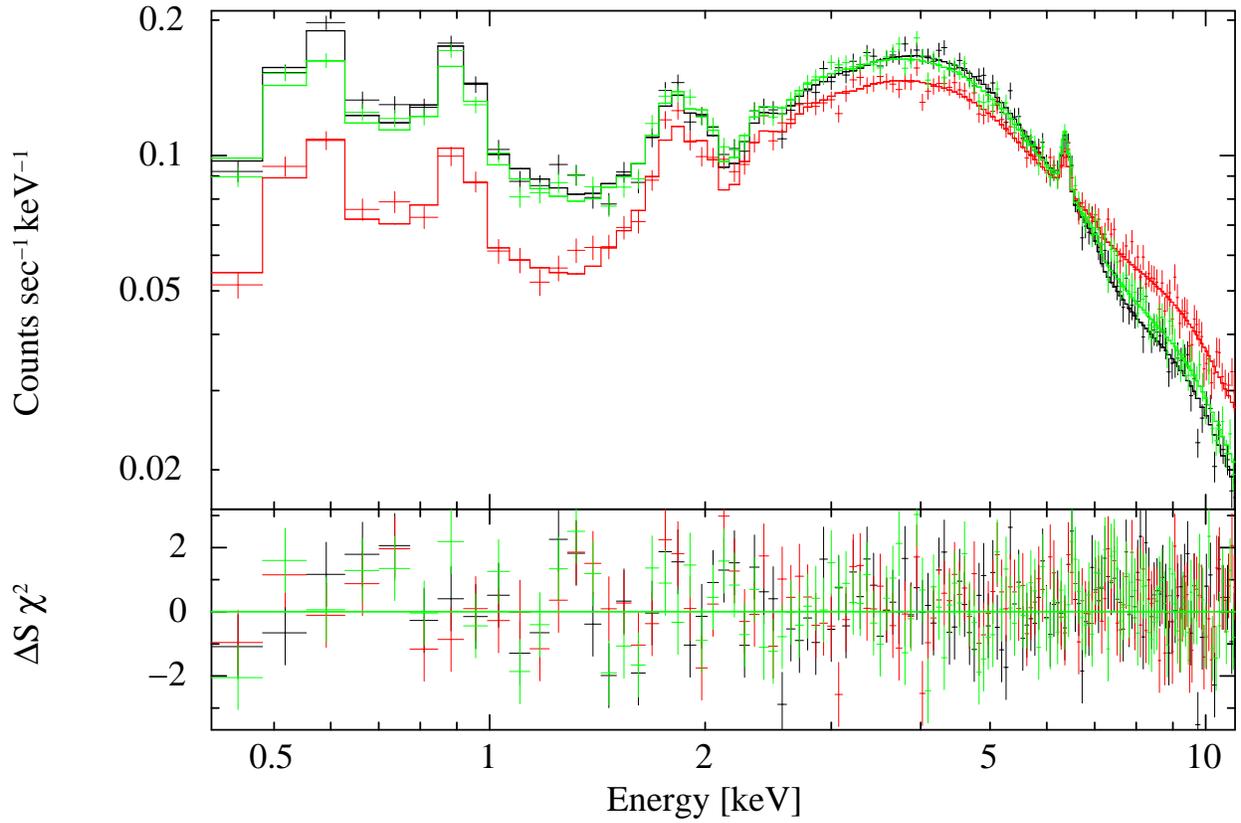}}
\caption{The spectral model with partial covering and a Compton reflection component, derived by fitting the three segments simultaneously.  See Table~\ref{pcfabs} for fitting parameters.\label{refabs}}
\end{figure}

\clearpage

\begin{deluxetable}{lcccccccccccc}
\rotate
\tabletypesize{\scriptsize}
\tablecaption{Power-Law Spectral Fitting with (A) PSF Core, (B) PSF Wing, and (C) Readout Streak\label{simple_tab}}
\tablewidth{0pt}\tablecolumns{13}
\tablehead{
\colhead{Fitting} & &
\multicolumn{3}{c}{Obs 9217a} & &
\multicolumn{3}{c}{Obs 9217b} & &
\multicolumn{3}{c}{Obs 9218}\\
\cline{3-5} \cline{7-9} \cline{11-13} 
\colhead{Parameter} & & \colhead{A} & \colhead{B} & \colhead{C} & &
\colhead{A} & \colhead{B} & \colhead{C}  & &
\colhead{A} & \colhead{B} & \colhead{C}
}
\startdata
PhoIndx $\Gamma_1$&& $2.06^{+0.15}_{-0.05}$& $3.12\pm 0.28$& $2.99\pm 0.44$&& $2.05^{+0.08}_{-0.07}$& $3.58_{-0.3}^{+0.3}$& $3.10_{-0.4}^{+0.13}$&& $2.09^{+0.15}_{-0.04}$& $3.19\pm 0.21$ & $2.67\pm 0.23$\\
Normalization $A_1\times 10^{-4}$&& $4.5^{+0.6}_{-0.9}$ & $20.0^{+2.9}_{-2.4}$ & $4.5^{+0.9}_{-1.4}$ &&$4.1^{+1.1}_{-0.2}$ &  $25.6^{+2.5}_{-2.3}$& $4.8\pm 1.2$ && $4.4^{+1.2}_{-0.9}$ & $22.5\pm 2.2$& $6.7^{+0.9}_{-1.2}$\\
Absorption $N_{H,2}$ ($10^{22}$ cm$^{-2}$)&& $3.52^{+0.16}_{-0.15}$& $2.82^{+0.61}_{-1.01}$& $3.00^{+0.69}_{-0.59}$ && $3.55^{+0.16}_{-0.17}$ & $2.50^{+0.63}_{-0.67}$& $2.58^{+1.03}_{-0.41}$&& $3.64^{+0.15}_{-0.16}$ & $3.42^{+0.96}_{-0.85}$& $3.56^{+1.20}_{-0.77}$\\
PhoIndx $\Gamma_2$&& $0.68\pm 0.05$& $0.43_{-0.37}^{+0.28}$& $0.74_{-0.05}^{+0.09}$&& $0.93\pm 0.08$& $0.47_{-0.09}^{+0.18}$& $0.86^{+0.06}_{-0.10}$&&$0.89\pm 0.07$ &$0.69_{-0.26}^{+0.28}$ & $0.77\pm 0.13$\\
Normalization $A_2\times 10^{-3}$&& $3.7^{+1.0}_{-0.9}$& $1.9^{+1.3}_{-0.7}$& $3.5^{+0.6}_{-0.5}$ && $5.0^{+2.8}_{-2.9}$ & $3.3^{+0.9}_{-0.6}$&  $5.1^{+2.5}_{-1.9}$ && $3.7^{+1.0}_{-0.9}$ & $3.5^{+2.1}_{-1.2}$ & $5.8^{+1.3}_{-1.9}$\\
Flux $F_{2-10{\rm{keV}}}$ ($\times 10^{-11}$erg s$^{-1}$ cm$^{-2}$)&& $8.9^{+1.8}_{-2.5}$ & $7.6^{+4.6}_{-2.8}$& $6.1^{+0.6}_{-2.7}$&& $10.0_{-3.0}^{+1.2}$ & $11.3_{-2.2}^{+1.8}$& $8.4\pm 1.2$&& $8.6^{+2.7}_{-2.9}$& $8.1^{+3.0}_{-2.5}$ & $9.0^{+0.8}_{-3.9}$\\
Goodness of Fit ($\chi^2/d.o.f$)&& 641/664 & 96/93 & 44/46 && 614/705 & 133/127 & 97/69 && 619/679 & 123/114 & 55/71 \\
Pileup Fraction && 32\%& ... & ... && 41\%& ... & ... && 35\%& ... & ... \\
%&& & & && & & && & & \\
\enddata \tablecomments{A fixed Galactic column $N_{H,1}=2\times
  10^{20}$  cm$^{-2}$ is assumed for the first absorption column (Yang et
  al. 2001). $\Gamma_1$ and $A_1$ is the photon index and the
  normalization for the soft power law component,
  respectively. $N_{H,2}$, $\Gamma_2$, and $A_2$ is the absorption
  column, the photon index, and the normalization for the hard power
  law component, respectively.}
\end{deluxetable}

\clearpage
\pagestyle{empty}
\begin{deluxetable}{cccccccc}
\rotate
\tabletypesize{\scriptsize}
\tablecaption{Reflection Fitting with Pileup Model\label{reflect_tab}}
\tablewidth{0pt}
\tablehead{
\colhead{\begin{tabular}{c}
OBSID\\
(1)
\end{tabular}} &
\colhead{\begin{tabular}{c}
Absorption\\
$N_{H,1}$\\
($10^{22}$ cm$^{-2}$)\\
(2)
\end{tabular}} &
\colhead{\begin{tabular}{c}
PhoIndx\\
$\Gamma_1$\\
(3)
\end{tabular}} &

\colhead{\begin{tabular}{c}
Fe Flux\\
($10^{-5}$photon s$^{-1}$ cm$^{-2}$)\\
(4)
\end{tabular}} &
\colhead{\begin{tabular}{c}
Fe EW\\
(eV)\\
(5)
\end{tabular}} &
\colhead{\begin{tabular}{c}
$F_{2-10{\rm{keV}}}$\\
($10^{-11}$erg $s^{-1}$ cm$^{-2}$)\\
(6)
\end{tabular}} &
\colhead{\begin{tabular}{c}
Refl. Factor\\
$R=\Delta\Omega/2\pi$\\
(7)
\end{tabular}} &
\colhead{\begin{tabular}{c}
Pile-Up\\
Fraction\\
(8)
\end{tabular}}
}
\startdata
&&& (A) & $\chi^2/d.o.f=2259/2033$ &&&\\
\hline\\
9217a & $0.02^{+0.01}_{-0.02}$ & $2.18\pm0.15$  &$6.2^{+5.4}_{-2.0}$ & $86$ & $5.3^{+0.2}_{-1.0}$ & $3.5\pm 1.0$ & 35\% \\
9217b & $0.04\pm0.02$        & $2.30\pm0.13$  &$9.2^{+14.}_{-3.0}$ & $77$ & $9.8^{+0.1}_{-1.0}$ & $1.6\pm 0.3$ & 42\% \\
9218 & $0.02^{+0.02}_{-0.01}$  & $2.29^{+0.07}_{-0.09}$ & $10.9^{+2.7}_{-2.5}$ & $113$ & $6.0^{+0.2}_{-1.2}$ & $3.0\pm 0.7$ & 37\% \\
\hline\\
&&& (B) & $\chi^2/d.o.f=2380/2033$ &&&\\
\hline\\
9217a & $0.27\pm0.04$ & $4.54\pm0.4$ &$15\pm 4$ & $86$ & $7.7^{+0.2}_{-1.0}$ &  $4.3\pm 0.4$ & 35\% \\
9217b & $0.30\pm0.02$ & $4.77\pm0.13$ &$13\pm5$ & $64$ & $9.6^{+0.1}_{-1.0}$ & $3.5\pm 0.3$ & 42\% \\
9218 & $0.28^{+0.02}_{-0.03}$  & $4.56\pm0.2$ & $14\pm6$ & $99$ & $8.0^{+0.2}_{-1.2}$ & $3.6\pm 0.3$ & 37\% \\
\enddata

\tablecomments{Model (A) allows zero width gaussian emission lines to vary and Model (B) uses gaussian lines frozen to the values measured in the HETG spectra (Ogle et al. 2000).} 

\tablecomments{$N_{H,1}$ and $\Gamma_1$ are the absorption column and the photon index for the soft power law component, respectively.  The following parameters were fitted simultaneously for all three spectra: (A) $N_{H,2}=4.05\pm 0.16\times 10^{22}$cm$^{-2}$, $\Gamma_2=\Gamma_{PEXRAV}=1.15^{+0.08}_{-0.17}$; (B) $N_{H,2}=3.62\pm0.14\times 10^{22}$cm$^{-2}$, $\Gamma_2=\Gamma_{PEXRAV}=1.09\pm0.09$, where $N_{H,2}$, $\Gamma_2$ and $\Gamma_{PEXRAV}$ are the absorption column for the hard power law component, the photon index for the hard power law component and for the Compton reflection component, respectively; normalization for the Compton reflection component: (A) $A_{PEXRAV}=1.81\pm0.08\times 10^{-2}$;  (B) $A_{PEXRAV}=3.78\pm 0.15\times 10^{-2}$.}
\end{deluxetable}

\clearpage
\pagestyle{empty}
\begin{deluxetable}{ccccccccc}
\rotate
\tabletypesize{\scriptsize}
\tablecaption{Partially Covered Reflection Fitting with Pileup Model\label{pcfabs}}
\tablewidth{0pt}
\tablehead{
\colhead{\begin{tabular}{c}
OBSID\\
(1)
\end{tabular}} &
\colhead{\begin{tabular}{c}
Absorption\\
$N_{H,1}$\\
($10^{22}$ cm$^{-2}$)\\
(2)
\end{tabular}} &
\colhead{\begin{tabular}{c}
PhoIndx\\
$\Gamma_1$\\
(3)
\end{tabular}} &
\colhead{\begin{tabular}{c}
Fe Flux\\
($10^{-5}$photon s$^{-1}$ cm$^{-2}$)\\
(4)
\end{tabular}} &
\colhead{\begin{tabular}{c}
Fe EW\\
(eV)\\
(5)
\end{tabular}} &
\colhead{\begin{tabular}{c}
$F_{2-10{\rm{keV}}}$\\
($10^{-11}$erg $s^{-1}$ cm$^{-2}$)\\
(6)
\end{tabular}} &
\colhead{\begin{tabular}{c}
Covering \\
Fraction\\
(7)
\end{tabular}} &
\colhead{\begin{tabular}{c}
Refl. Factor\\
$R=\Omega/2\pi$\\
(8)
\end{tabular}} &
\colhead{\begin{tabular}{c}
Pileup\\
Fraction\\
(9)
\end{tabular}}
}
\startdata
&&&& (A) &$\chi^2/d.o.f=2219/2029$&&&\\
\hline
9217a & $0.12\pm0.03$ & $3.06^{+0.08}_{-0.10}$ & $8.1^{+2.0}_{-1.9}$ & $98$ & $6.9\pm 0.2$ & $0.47\pm0.07$ & $2.3\pm0.2$ & 35\% \\
9217b & $0.10\pm0.04$& $2.86^{+0.09}_{-0.14}$ & $9.7^{+2.8}_{-3.6}$ & $86$ & $9.8^{+0.1}_{-0.2}$ & $0.35\pm0.09$ & $1.5\pm0.1$ & 42\% \\
9218 &  $0.09\pm0.04$ & $2.98^{+0.07}_{-0.10}$ & $8.3^{+1.8}_{-1.7}$ & $115$ & $6.0^{+0.1}_{-0.3}$ & $0.35\pm0.09$ & $2.7\pm0.2$ & 39\% \\
\hline\\
&&&& (B) &$\chi^2/d.o.f=2371/2029$&&&\\
\hline\\
9217a & $0.19\pm0.05$ & $4.1\pm0.2$ & $18.8\pm9.5$ & $94$ & $7.7^{+0.2}_{-0.3}$ & $0.42\pm0.07$ & $4.5\pm1.1$ & 35\% \\
9217b & $0.50\pm0.07$& $6.3\pm0.5$ & $9.5\pm9.0$ & $64$ & $9.6^{+0.1}_{-0.5}$ & $0.16\pm0.11$ & $6.8\pm1.7$ & 42\% \\
9218 &  $0.38\pm0.05$ & $5.5\pm0.4$ & $16.1\pm9.0$ & $105$ & $8.2^{+0.1}_{-0.4}$ & $0.22\pm0.10$ & $6.5\pm1.5$ & 39\% \\
\enddata

\tablecomments{Model (A) allows zero width gaussian emission lines to vary and Model (B) uses gaussian lines frozen to the values measured in the HETG spectra (Ogle et al. 2000).} 

\tablecomments{$N_{H,1}$ and $\Gamma_1$ are the absorption column and
  the photon index for the soft power law component, respectively. The
  following parameters were fitted simultaneously for all three
  spectra: (A) $N_{H,2}=4.3\pm 0.3\times 10^{22}$ cm$^{-2}$,
  $\Gamma_2=\Gamma_{PEXRAV}=1.68^{+0.04}_{-0.15}$, and
  $N_{H,PCFABS}=1.9\pm 0.2\times 10^{23}$ cm$^{-2}$; (B)
  $N_{H,2}=4.4\pm 0.3\times 10^{22}$ cm$^{-2}$,
  $\Gamma_2=\Gamma_{PEXRAV}=1.67\pm0.16$, and
  $N_{H,PCFABS}=1.3\pm0.3\times 10^{23}$ cm$^{-2}$, where
  $N_{H,PCFABS}$ is the column for the partial covering absorber,
  $N_{H,2}$ and $\Gamma_2$ are the absorption column and the photon
  index for the hard power law component, respectively; normalization
  for the Compton reflection component: (A) $A_{PEXRAV}=3.37\pm
  0.18\times 10^{-2}$; (B) $A_{PEXRAV}=0.13\pm 0.04$.}

\end{deluxetable}

\begin{deluxetable}{ccccccccc}
\rotate
\tabletypesize{\scriptsize}
\tablecaption{Ionized Absorber Fitting with Pileup Model\label{absori_tab}}
\tablewidth{0pt}
\tablehead{
\colhead{\begin{tabular}{c}
OBSID\\
(1)
\end{tabular}} &
\colhead{\begin{tabular}{c}
Absorption\\
$N_{H,1}$\\
($10^{22}$ cm$^{-2}$)\\
(2)
\end{tabular}} &
\colhead{\begin{tabular}{c}
PhoIndx\\
$\Gamma_1$\\
(3)
\end{tabular}} &

\colhead{\begin{tabular}{c}
Fe Flux\\
($10^{-5}$photon s$^{-1}$ cm$^{-2}$)\\
(4)
\end{tabular}} &
\colhead{\begin{tabular}{c}
Fe EW\\
(eV)\\
(5)
\end{tabular}} &
\colhead{\begin{tabular}{c}
$F_{2-10{\rm{keV}}}$\\
($10^{-11}$erg $s^{-1}$ cm$^{-2}$)\\
(6)
\end{tabular}} &
\colhead{\begin{tabular}{c}
Ionization\\
$\xi$\\
(7)
\end{tabular}} &
\colhead{\begin{tabular}{c}
Reflection\\
$R$\\
(8)
\end{tabular}} &
\colhead{\begin{tabular}{c}
Pile-Up\\
Fraction\\
(9)
\end{tabular}}
}
\startdata
9217a & $0.26^{+0.03}_{-0.02}$ & $4.7\pm0.4$  &$14.4^{+5.4}_{-3.0}$ & $89$ & $7.7^{+0.2}_{-1.6}$ & $6.5\pm0.4$& $9.7\pm 1.0$ & 35\% \\
9217b & $0.56\pm0.08$        & $6.3\pm0.6$  &$10.4^{+9.6}_{-4.0}$ & $62$ & $9.6^{+0.1}_{-1.0}$ & $13.7^{+0.6}_{-0.3}$& $9.0\pm 0.8$ & 42\% \\
9218 & $0.45\pm0.05$  & $5.7\pm0.5$ & $16.1^{+8.2}_{-5.5}$ & $99$ & $8.2^{+0.2}_{-1.6}$ & $14.0^{+0.5}_{-0.3}$ & $9.5\pm 1.0$ & 37\% \\
\enddata
\tablecomments{Model uses gaussian lines frozen to the values measured in the HETG spectra (Ogle et al. 2000). $N_{H,1}$ and $\Gamma_1$ are the absorption column and the photon index for the soft power law component, respectively.  The following parameters were fitted simultaneously for all three spectra: $N_{H,2}=6.47\pm0.13\times 10^{22}$cm$^{-2}$, $\Gamma_2=1.65$ (frozen) where $N_{H,2}$, $\Gamma_2$ are the absorption column for the hard power law component, and the photon index for the hard power law component,  respectively. Goodness of fit $\chi^2/d.o.f=2429/2031$.}
\end{deluxetable}

\end{document}